\documentclass[journal]{IEEEtran}
\IEEEoverridecommandlockouts
\usepackage{etex}
\usepackage{amsmath,amssymb}
\usepackage{float}
\usepackage{lipsum}
\usepackage{array}
\usepackage{cite}
\usepackage{textcomp}
\usepackage{hhline}
\usepackage{siunitx}
\usepackage{balance}
\usepackage{makecell}
\usepackage{mathtools}
\usepackage{float}
\usepackage[pdftex]{graphicx}
\usepackage{siunitx}
\usepackage[dvipsnames]{xcolor}
\usepackage{xcolor}
\usepackage{colortbl}
\usepackage{tikz}
\usetikzlibrary{decorations}
\usetikzlibrary{calc} 
\tikzset{>=latex}
\usetikzlibrary{arrows,shapes}
\usetikzlibrary{arrows}
\usetikzlibrary{plotmarks}
\usepackage{ctable}
\usepackage{pgfplots}
\usepackage[english]{babel}
\usepackage{color}
\usepackage[utf8]{inputenc}
\usepackage[T1]{fontenc}
\usetikzlibrary{decorations.pathreplacing,shapes.misc}
\usetikzlibrary{patterns}
\usepackage{tcolorbox}
\usepackage{balance}
\usepackage{subfigure}
\usepackage{hyphenat}
\usepackage{amsmath,bm}
\usepackage[T1]{fontenc}
\usepackage{algpseudocode}
\usepackage[ruled,vlined]{algorithm2e}
\usepackage{graphicx}
\usepackage{subfig}

\addto\captionsenglish{}
\newcommand\scalemath[2]{\scalebox{#1}{\mbox{\ensuremath{\displaystyle #2}}}}
\newtheorem{prop}{Proposition}
\newtheorem{Ex}{Example}

\def\forcemath#1{\ifmmode #1 \else $#1$\fi}

\newcommand{\ASH}{\textcolor{black}}

\newcommand{\commentA}[2]{{\color{black}{#1}}}
\newcommand{\commentJ}[2]{{\color{black}{#2}}}

\newcommand{\commentAFF}[2]{{\color{black}{#2}}}

\definecolor{blue-violet}{rgb}{0.54, 0.17, 0.89}

\newcommand{\R}{\textcolor{black}}
\newcommand{\RR}{\textcolor{black}}
\newcommand{\RRR}{\textcolor{black}}

\newcommand{\Df}[0]{\Delta_f}
\newcommand{\Dw}[0]{\Delta_\omega}

\ifCLASSINFOpdf
\else
\fi

\begin{document}
	
	\title{Phase-based Ranging in Narrowband Systems  with Missing/Interfered Tones}

	\author{Alireza Sheikh, Jac Romme, Jochem Govers, Amirashkan Farsaei, and Christian Bachmann \\ \IEEEauthorblockA{
			\thanks{A. Sheikh, J. Romme, J. Govers, A. A. Farsaei, and C. Bachmann are with \commentJ{signal processing group in}{} imec, Holst Centre, High Tech Campus 31, 5656 AE Eindhoven, The Netherlands (email: alireza.sheikh@imec.nl).}
	}}
	
	
	\maketitle
	
	%\GL{The proposed algorithm closes over $50$\% of the performance gap
		%between iBDD and iterative decoding based on more complex soft-input
		%soft-output component-code decoders}.  
	
	\begin{abstract}
		The growth \commentJ{on}{in} the number of \commentJ{cheap}{low-cost} narrow band \commentJ{systems}{radios}  such as Bluetooth low energy (BLE) enabled applications such as asset tracking, human behavior monitoring, and keyless entry. The accurate range estimation is a must in such applications. Phase-based ranging has recently gained momentum due to its high accuracy in \commentJ{multiupath}{multipath} environment compared to traditional schemes such as ranging based on received signal strength. The phase-based ranging requires tone exchange on multiple frequencies \commentJ{in uniformly step-sized frequency grid}{on a uniformly sampled frequency grid}. Such tone exchange may not be possible due to some missing tones, e.g., \commentJ{reserved tones for advertisement in BLE}{reserved advertisement channels}. Furthermore, the IQ values at a given tone may be distorted by interference. In this paper, we proposed two phase-based ranging schemes which deal with the missing/interfered tones. \commentAFF{Using simulations, complexity analysis, and three measurement setups, we compare the performance and complexity of the proposed schemes.}{We compare the performance and complexity of the proposed schemes using simulations, complexity analysis and measurement setups}. In particular, we show that for small number of missing/interfered tones, the proposed system based on employing a trained neural network (NN) performs very close to a reference ranging system where there is no missing/interference tones. Interestingly, this high performance is at the cost of negligible additional computational complexity and up to $60.5$ Kbytes of additional required memory compared to the reference system, making it an attractive solution for ranging using hardware-limited radios such as BLE. 
		
	\end{abstract}

	
	%As an example, for a PC with double-error-correcting
	%Bose-Chaudhuri-Hocquenghem component codes, the net coding gain can be
	%increased by up to $0.23$--$0.57$ dBs.  
	
	\IEEEpeerreviewmaketitle
	
	%\tableofcontents
	
	\section{Introduction}\label{Sec:Intro}
	
	\IEEEPARstart{T}{he} group of connected devices, often referred to as internet of things, is widely available nowadays and the number of IoT devices will exceed $29$ billions devices by $2030$ \cite{IoT2022}. This growth is due to availability of cheap narrow-band systems such Bluetooth low-energy (BLE) or Zigbee, as they are widely available in smart phones, airports, shopping malls, and factories. The IoT systems can be used in many applications such as asset tracking, human behavior monitoring used for smart elderly care \cite{BARSOCCHI201526}, and key-less entry. The main required functionality for such applications is the distance estimation between two devices, which in turn can be used for localization.  Although it is known that ultra wide band (UWB) systems provide better ranging accuracy compared to the narrow-band systems, its usage is rather limited in IoT systems due to its \commentJ{costs}{cost} \cite{Lu2021}.
	
	The distance between two radio systems can be estimated based on three main approaches: (i) Received signal strength indicator (RSSI), (ii) \R{Time-of-flight (ToF)} or Time-difference-of-arrival (TDoA), (iii) phase-based ranging. RSSI-based ranging suffers from low accuracy \cite{Wisanmongkol2019}, \R{\cite{Giovanelli2018}} as in \commentAFF{a real environment}{practice,} multipath violates the direct relation of distance to RSSI. \R{ToF} and TDoA are also effective in the UWB system, where the bandwidth and hence, time resolution is high \cite{Shen2010}. In narrow-band systems, phase-based ranging is a popular candidate~\cite{Kluge2013}-\cite{Zand2019}. \R{A comprehensive overview of accurate localization systems can be found in \cite{CHANG2022}.}
	
	%as \commentJ{it sequentially measures the frequency response of multiple narrowband channels and combines them to improve the ranging accuracy,}{} \commentJ{which is known as multi-carrier phase difference (MCPD) ranging}{\it empty} 
	%\comment{virtual}{why virtual? We actually measure it. Virtual is often in another context, where N "observations" are created from M measurements with M<N  (e.g. Virtual antenna arrays). } higher bandwidth, 

	In phase-based ranging, the \commentJ{phase or IQ values are}{channel is} measured between two devices (known as initiator and reflector) \commentJ{on multiple tones in }{on a uniform frequency grid over} the bandwidth of interest \cite{Zand2019}. The \commentJ{phase}{} measurements can be processed using fast Fourier transform (FFT) \cite{Yuan2018} \commentJ{}{to obtain the CIR}, however, the ranging \commentJ{performance}{accuracy} is not great in \commentJ{multiupath}{multipath} environment due to limited resolution of the FFT \cite{boer2019}. \commentJ{Using multiple IQ measurements on the frequency band, the}{Instead of FFT,} super-resolution algorithms such as Multiple signal classification (MUSIC) can be used to improve the accuracy of ranging in \commentJ{multiupath}{multipath} environments \cite{Miesen2012}. \R{In \cite{boer2019} a novel ranging scheme based on single-snap shot MUSIC \cite{Liao2014MUSICFS} is proposed which only employs the IQ measurements over the frequency band.} \R{In \cite{boer2019}, it is shown that range estimation using the reconstructed channel as a one-way channel response is more accurate compared to range estimation using the two-way channel response which does not require the channel reconstruction.} 
	
	The critical assumption of applying the super-resolution algorithms such as MUSIC on phase-based ranging is that the IQ measurements at the initiator and reflector are \commentJ{available from a uniformly step-sized frequency }{taken on a uniform, frequency domain sampling} grid \cite{boer2019}. However, this assumption will be violated in the case that there are missing or interfered tones. Missing tones in Bluetooth system are unavoidable as some \commentJ{tones}{channels} are reserved for Bluetooth advertising packets \cite{Gomez_2012}. Therefore, the initiator and reflector cannot measure \commentJ{IQ on such}{on these} advertisement \commentJ{tones}{channels}. Furthermore, due to \commentJ{the presence}{interference} of other systems such as Bluetooth, Zigbee, or WIFI (which uses the same bandwidth), \commentJ{ the IQ values at the interfered tones}{some measurements} cannot be trusted. To the best of \commentJ{ }{our} knowledge, there is no work in literature which evaluates the effect of missing/interfered tones on accuracy of phase-based ranging in \commentJ{multiupath}{multipath} environments. \R{We highlight that the contribution of this paper is mainly about how to deal with the missing/interfered tones in phase-based ranging based on super resolution algorithm. To this end, we used MUSIC as a signal processing tool for evaluation of our methods. There are other signal-noise subspace separation methods that can be used, e.g., ESPRIT \cite{RoyESPRIT}, MVDR \cite{MVDR1969}. It is shown that MUSIC performs better than MVDR and ESPRIT in the direction of arrival estimation problem \cite{Akbari2010}, \cite{Oumar2012}.}
	
	%We highlight that in this paper, we use MUSIC as one of the signal-noise subspace separation methods to estimate the range.  
	
	\R{There exists prior-art on the performance of the MUSIC algorithm using a non-uniform sampling grid. For example, MUSIC-based direction of arrival estimation is possible when using  sparse, non-uniform antenna arrays, see e.g. \cite{Abramovich98,shakeri2012direction}. These works however  assume availability of multiple uncorrelated snapshots, and do not require any channel reconstruction (see section \ref{MUSIC_general} for channel reconstruction). Other works analyze direction of arrival estimation with MUSIC using randomly missing observations \cite{Suryaprakash2015,setayesh2019direction}, but they rely on many snapshots with uncorrelated sources, knowledge of the number of sources, and additionally do not require channel reconstruction, which makes them not applicable to our problem statement.}
	
	In this paper, we propose \commentA{two}{or three?} schemes for phase-based ranging using MUSIC as a super resolution algorithm in various \commentJ{multiupath}{multipath} environments, where some tones are missing/interfered. In the first scheme, the IQ values corresponding to missing/interfered tones are neglected, and the cost function of the MUSIC algorithm is modified to infuse the information of the remaining IQ values. In the second scheme, we first estimate the squared frequency response of the channel at missing/interfered tones, and then apply the MUSIC algorithm to estimate the range. The estimation of the squared frequency response of the channel at missing/interfered tones are done \commentJ{}{either} using atomic norm minimization \commentJ{and}{or} using a trained NN. The performance of \commentJ{all}{both} schemes \commentJ{are}{is} compared via simulations and further validated using measurement setups. Based on the complexity analysis of \commentJ{all}{both} schemes and observed performance, we show that the proposed system based on NN provides the best performance-complexity trade-off. \R{We highlight that the conclusions drawn based on the second scheme (recovering the missing/interfered tones) can be also drawn if other super resolution algorithms are used for the range estimation.}
	
	The remainder of the paper is organized as follows\commentJ{.}{:} In Section~\ref{Sec:Prel}, some preliminaries are explained and ranging based on \cite{boer2019} \commentJ{are}{is} reviewed. The proposed schemes for phase-based ranging with missing/interfered tones are discussed in section~\ref{Sec:ranging_mis_interf}. In Section~\ref{Sec:Comp_complexity}, we compare the complexity of the proposed schemes. The simulation and measurement results are given in \commentJ{section}{Section}~\ref{Sec:Sim}. Finally, conclusions are drawn in Section~\ref{Sec:Conclusions}.
	
	\textbf{Notation}: We use boldface letters for vectors
	and matrices, e.g., $\boldsymbol{x}$  and $\boldsymbol{X}$. Furthermore, $\boldsymbol{x}^2$ stands for the element-wise squared of $\boldsymbol{x}$. $(\cdot)^\mathsf{T}$ and $(\cdot)^\mathsf{H}$ are the matrix transpose and Hermitian operations, respectively. $c$ denotes the speed of light and $\boldsymbol{I}_L$ is the identity matrix of size $L \times L$. $\commentJ{Teop}{Toep}(\boldsymbol{x})$ is a Toeplitz matrix \cite{Petersen2008} whose first column is equal to $\boldsymbol{x}$. $trace(\boldsymbol{X})$ is a function returning the sum of the main diagonal components of $\boldsymbol{X}$. $\boldsymbol{X} \succ 0$ means that $\boldsymbol{X}$ is a positive definite matrix. $\measuredangle$ denotes the unwrapped angle. $\leq$ stands for less than or equal to. \ASH{$\mathcal{O}(.)$ stands for the order of the complexity}. $Real(.)$ and $Imag(.)$ denotes the real and imaginary components of a complex value. Slice over integer numbers are shown using ``$:$'', e.g., $\{1:4\}$ and $\{1,2,3,4\}$ are equivalent. $\{\}$ stands for the empty set. $A$ $\leftarrow$ $B$ means that $A$ is overwritten by $B$. Finally, ``$\propto$`` and ``$==$`` denote \commentJ{``equal to''}{proportional and a logical comparison, respectively}. 
%	$\text{diag}(\boldsymbol{X})$ gives a vector comprising the main diagonal of $\boldsymbol{X}$. Furthermore, $\text{diag}(\boldsymbol{x})$ gives a square matrix, where $\boldsymbol{x}$ is placed as the diagonal components of such matrix.

	\section{Preliminaries}\label{Sec:Prel}
	
	In \commentAFF{the following section, we first explain some preliminaries}{this section, we first present the system model}. 
	%IQ measurements at the initiator and reflector and the frequency channel response corresponding to such IQ measurements. 
	Then, we briefly explain the structure of phase-based ranging based on IQ measurements using super resolution algorithm such as MUSIC, which was introduced in \cite{boer2019}.  
	
	\subsection{\commentJ{Frequency channel model}{System Model}}\label{Sec:SysMod}
	
\begin{figure}[t] \centering 
		\includegraphics[scale=0.88]{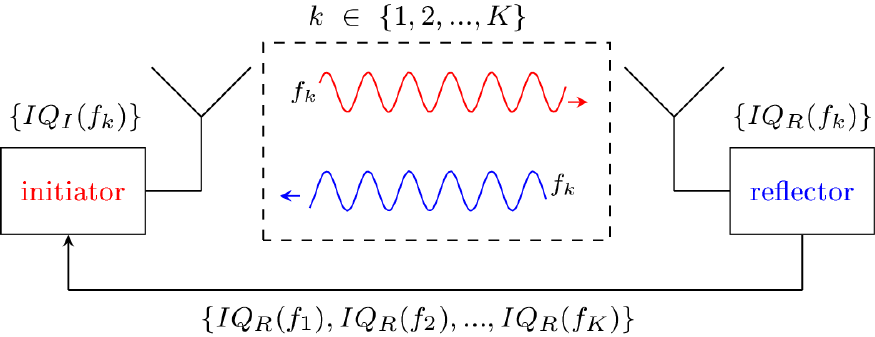}  
		\vspace{-1ex}
		\caption{\RR{Schematic of the phase-based ranging data capture.}}  \vspace{-1ex}
		\label{Init_reflec_IQ} 
	\end{figure}

	\commentJ{
		Let us assume that the coherence bandwidth of the wireless channel is larger than the overall bandwidth of the narrow-band system\footnote{As an example, the overall bandwidth of Bluetooth system in the 2.4 GHz ISM band is 80 MHz.}. This means that the channel will not vary during the IQ measurements. We also assume that the available bandwidth of the narrow band system is swiped with the constant frequency step size of $\Df$. We consider a sequential channel measurements between initiator and reflector. First the initiator sends a tone to the reflector and then reflector measures the IQ. Then, the reflector sends the same tone and initiator measures the corresponding IQ. Afterwards, both initiator and reflector switch to the next frequency and perform the IQ measurement for a new tone. This procedure continues for the entire frequency band.
		The frequency response of the channels  corresponding to the $k$-th tone is given as
		\begin{equation}\label{eq:Chanmodel}
			\scalemath{0.93}{{h_k} = \sum\limits_{m = 0}^{M - 1} {{a_m}} \exp ( - j{\omega _0}{\tau _m})\exp ( - jk\Dw {\tau _m}),k \in \{ 1,...,K\} },
		\end{equation}
		where $\omega_0=2\pi f_0$, $f_0$ is the carrier frequency, $\Dw = 2 \pi \Df$, $M$ is the number of paths, $a_m$ is the coefficient of $m$-th path, and $\tau_m$ is the delay corresponding to $m$-th path. The delay corresponding to line-of-sight (LoS) path is $\tau_0$.
		
	}{
		During phase-based ranging, two narrowband radios collaborate to measure the frequency response of the propagation channel between them. First the so-called initiator sends an unmodulated tone on the first channel to the other radio, called reflector, such that it can measures the channel response. After given time, the reflector sends a tone on the same channel and the initiator will measure the channel response. Then, both radio switch to the next channel and the procedure is repeated again, until the channel response is measured over bandwidth $B$ on a uniform frequency grid with step size $\Df$. \RR{As will be explain in the following, during this procedure, both initiator and reflector measure the IQ values. Then, the IQ values of reflector are sent to the initiator, where the range will be estimated. This procedure is schematically shown in Fig.~\ref{Init_reflec_IQ}.}
		
		Assuming the whole procedure is conducted within the coherence-time of the channel, its response to a tone with frequency $f_k$ can be modeled as
		\begin{equation}\label{eq:Chanmodel}
			\scalemath{0.93}{{h_k} = \sum\limits_{m = 0}^{M - 1} {{a_m}} \exp ( - j{\omega _0}{\tau _m})\exp ( - jk\Dw {\tau _m}),k \in \{ 1,\hdots,K\} },
		\end{equation}
		where $\omega_0=2\pi f_0$, $f_0$ is the carrier frequency, $\Dw = 2 \pi \Df$, $M$ is the number of paths, $K$ is the number of tones, $a_m$ is the coefficient of $m$-th path, and $\tau_m$ is the delay corresponding to $m$-th path. The delay corresponding to line-of-sight (LoS) path is $\tau_0$.
		
	}

	%\comment{\subsection{System Model}\label{Sec:SysMod}}{}

	%We consider a sequential channel measurements between initiator and reflector. First the initiator sends first tone to the reflector and then reflector measures the IQ. Then, the reflector sends the same tone and initiator measures the corresponding IQ. Then, both initiator and reflector switch to the next frequency incoherently and perform the IQ measurement for a new tone. This procedure continues for the entire available frequency band. 

	\begin{figure}[t] \centering 
		\includegraphics[scale=0.78]{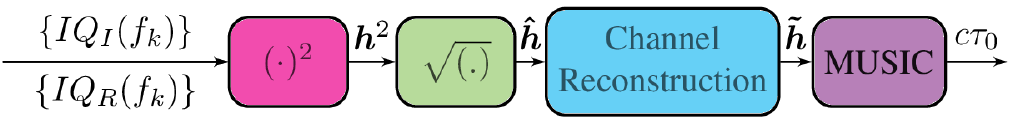}  
		\vspace{-1ex}
		\caption{Schematic of the phase-based ranging using super resolution algorithm.}  \vspace{-3ex}
		\label{Ranging_Super} 
	\end{figure} 
	
	Similar to \cite{yao2021ranging}, \cite{boer2019}, \commentJ{the Phase Locked Loop (PLL)}{we assume both devices have a homodyne receiver architecture, where the local-oscillator (LO) signal is used to both generate tones and as a mixer signal. The LO signal is typically generated using a Phase Locked Loop (PLL), where} in both initiator and reflector does not necessarily remain in lock when switching between different \commentJ{tones}{channels}, which is a more commercialized PLL compared to phase coherent counterpart \cite{Bechthum2020}. \commentJ{moved:}{However, we do assume that the LO of both initiator and reflector is phase-continuous when switching its role from transmitting to receiving mode or vice-versa.} The noise-free IQ samples corresponding to $k$-th tone measured at the reflector \RRR{($I{Q_R}({f_k})$)} and initiator \RRR{($I{Q_I}({f_k})$)} can be modeled as
	\begin{align}\label{eq:IQ}
		I{Q_R}({f_k}) = {A_R e^{{\theta _k}}}{h_k}, I{Q_I}({f_k}) = {A_I e^{ - {\theta _k}}}{h_k},
	\end{align}
	where $\theta _k$ is the uniformly random phase difference between initiator and reflector PLLs. \commentJ{We also assume that the local oscillator of both initiator and reflector keeps running when switching role from transmitting to receiving mode.Therefore, such}{As we assumed phase-continuous LO signals, the}  phase difference of LOs manifests as a phase with different signs in \eqref{eq:IQ}. \commentJ{\it added:}{The variables $A_R$ and $A_I$ model the TX power and RX gain of the radios combined. It is important to emphasize that we assume these to be constant for all channels. \RRR{In the following sections, we assume that the IQ values at the initiator and reflectors are normalized based on the $A_R$ and $A_I$ values, therefore, one can assume $A_R = A_I = 1$ in \eqref{eq:IQ}.}} \commentA{}{"As we assumed phase-continuous LO signals, the phase difference manifests as a phase with different signs in (2)." does it mean "the phase difference between the two LOs"}

	\subsection{\commentJ{\it added:}{Signal processing chain}}\label{MUSIC_general}
	
	The schematic of phase-based ranging using MUSIC as super-resolution algorithm is shown in Fig.~\ref{Ranging_Super}. First the two-way channel response of the $k$-th tone is computed by simply multiplying the IQ samples, i.e., \RRR{$h_k^2 = I{Q_R}({f_k})I{Q_I}({f_k})$}. As shown in \cite{boer2019}, the ranging accuracy based on the two-way channel responses is lower than ranging based on the one-way channel response in multipath channel. Therefore, \commentAFF{}{we use} the one-way channel response for all tones\commentAFF{is}{, which is} computed as
	\begin{equation}\label{eq:TWCR}
		{\hat{h}_k} \propto \sqrt {I{Q_R}({f_k})I{Q_I}({f_k})}  \commentJ{=}{\propto} {c_k}{h_k},{c_k} \in \{  \pm 1\},
	\end{equation}
	where $\hat{h}_k$ is the estimated frequency of the $k$-th tone. Note that due to square root operation in \eqref{eq:TWCR}, there is a $\pi$ phase ambiguity between $\hat{h}_k$ and ${h}_k$. This phase ambiguity can be resolved by employing channel reconstruction method proposed in \cite{yao2021ranging}, which attempts to maintain the phase progression between tones\commentJ{}{, assuming that $\Df$ is smaller than the coherence band}. The resulting estimated frequency response after channel reconstruction for $k$-th tone is denoted by $\tilde{h}_k$. After channel reconstruction, the vector $\boldsymbol{\tilde{h}}=[\tilde{h}_1,\cdots,\tilde{h}_K]$ is used as the input for MUSIC to estimated the distance. Note that due to the uniform frequency step-size between different tones, there is a constant phase rotation from tone $k$ to tone $k+1$ for each path in \eqref{eq:Chanmodel}. This property \commentJ{can be exploited}{is used} in the single snapshot MUSIC \cite{Liao2014MUSICFS} to find the time delays. In particular, spatial smoothing parameter \cite{Tie1985} $L$ can be applied to $\boldsymbol{\tilde{h}}$ to compute the Hankel matrix $\boldsymbol{H}$ given as    
	\begin{equation}\label{eq:Hankel}
		\boldsymbol{H} = \left[ {\begin{array}{*{20}{c}}
				{{\tilde{h}_1}}& \cdots &{{\tilde{h}_{L}}}\\
				\vdots & \ddots & \vdots \\
				{{\tilde{h}_{K-L+1}}}& \cdots &{{\tilde{h}_K}}
		\end{array}} \right].
	\end{equation}
	Then, using eigenvalue decomposition of ${\boldsymbol{H}^H}\boldsymbol{H}$, the signal and noise sub-spaces are computed as
	%\begin{equation}\label{eq:SN}
	%	{\boldsymbol{H}^H}\boldsymbol{H} = \boldsymbol{V}\boldsymbol{\Lambda} {\boldsymbol{V}^H},\boldsymbol{V} = [{\boldsymbol{v}_1}, \cdots ,{{\boldsymbol{v}}_L}] = [{\boldsymbol{V}^S}{\boldsymbol{V}^N}],
	%\end{equation}  
	\begin{equation}\label{eq:SN}
		{\boldsymbol{H}^H}\boldsymbol{H} = \boldsymbol{V}\boldsymbol{\Lambda} {\boldsymbol{V}^H},\boldsymbol{V} = [{\boldsymbol{V}_S}, {\boldsymbol{V}_N}],
	\end{equation}
	where 
	%each column of $\boldsymbol{V}$ is a Eigen vector, 
	$\boldsymbol{\Lambda}$ is a diagonal matrix comprising the eigenvalues, and \ASH{${\boldsymbol{V}_S}$} and \ASH{${\boldsymbol{V}_N}$} contain eigenvectors corresponding to signal and noise sub-spaces, respectively. We remark that the signal-noise sub-space separation can be done by setting the threshold on the eigenvalues \cite{Petersen2008}.
	%Note that the number of Eigen values (Eigen vectors) in \ref{label} is $L$. %hence, the forward smoothing parameter is a design parameter. 
	Based on \eqref{eq:Chanmodel}, \eqref{eq:Hankel}, and the Vandermonde decomposition of Hankel matrix, the steering vectors given as 
	\begin{equation}\label{eq:steer}
		\boldsymbol{e}(\tau ) = {[1,\exp ( - j\Dw\tau ),...,\exp ( - j\Dw(L - 1)\tau )]^T}
	\end{equation}
	should be orthogonal to the noise sub-space for the values of $\tau_n$. Hence, the pseudo spectrum function given as 
	\begin{equation}\label{eq:PS}
		\scalemath{0.8}{J(\tau) \triangleq\frac{1}{{\boldsymbol{e}{{(\tau )}^H}{{({\boldsymbol{V}_N})}^H}({\boldsymbol{V}_N})\boldsymbol{e}(\tau )}} = \frac{1}{{L - \boldsymbol{e}{{(\tau )}^H}{{({\boldsymbol{V}_S})}^H}({\boldsymbol{V}_S})\boldsymbol{e}(\tau )}}}
	\end{equation}
	can be computed, where the second equality follows by the fact that ${{({\boldsymbol{V}_N)}}^H}({\boldsymbol{V}_N})=\boldsymbol{I}_L-{{({\boldsymbol{V}_S})}^H}({\boldsymbol{V}_S})$. 
	The time delay $\tau_0$ can be estimated by finding the first peak of $J(\tau)$ \cite{Liao2014MUSICFS}, hence, the \commentJ{ }{estimated} distance between initiator and reflector is $c\tau_0$.
	
	\section{Ranging with missing/interfered tones}\label{Sec:ranging_mis_interf}

	In this section, we propose \commentA{two schemes}{three or two?} to estimate the range between initiator and reflector of a narrow-band system, when some tones are missing/interfered. In this paper, we refer to a “tone gap” as a block of consecutive tones where \commentJ{they}{measurements} are missing or interfered. Furthermore, an “available tone band” refers to a block of consecutive tones, where the frequency response of the channel can be estimated using IQ measurements as discussed in section ~\ref{Sec:SysMod}. 
	%Missing tones in Bluetooth system are unavoidable as some tones are reserved for transmitting Bluetooth advertising packets \cite{Gomez_2012}. Therefore, the initiator and reflector cannot measure IQ on such advertisement tones. Furthermore, due to the presence of other systems such as Bluetooth, Zigbee, or WIFI which uses the same bandwidth, the estimated channel of some tones cannot be trusted. 
	In this section, we assume that the missing/interred tones are known. \R{As explained before, missing tones are imposed by the standard and are known by any standard complaint device.} 
	
	%\R{Interfered tones can be detected in several manners, e.g. using the constant envelop property of the tones~\cite{Ferrara1985}, energy detection scheme~\cite{Kim2008}, statistical measure technique~\cite{Saad2015}, or employing the machine learning techniques~\cite{Xiwen2019,Grimaldi2019}. A simple interference detection based on constant envelop property of exchanged tones are explained in the following. Consider the IQ measurement at the reflector for $k$-th tone (corresponding to frequency $f_k$.)\footnote{\R{Note that the same mechanism can be used at the initiator to detect the interfered tones.}} The reflector down mix the tone at frequency $f_k$ with its LO (with frequency $f_k$) and take multiple samples. Then, the samples are averaged, giving $IQ_R(f_k)$. In the absence of interference, the measured envelop is constant, i.e., $|IQ_R(f_k)| \propto {h_k}$. However, with interference the measured envelop for each sample varies, as the IQ envelope is time dependent $|IQ_R(f_k)(t)|=|e^{\theta _k}{h_k}+f(t)|$, where $f(t)$ is the time dependent interference and can be due to e.g., WiFi or other BLE/Zigbee systems. Therefore, a simple method to identify the interference in phase based ranging is evaluating the envelope variation of the multiple IQ samples before averaging. If the variation is larger than a threshold, then the IQ is marked as interfered. If the IQ of the either of initiator or reflector is marked as interfered, such tone is considered as interfered.} 
	
	\R{Interfered tones can be detected in several manners, e.g. using the constant envelop property of the tones~\cite{Ferrara1985,jun2005}, energy detection scheme~\cite{Kim2008}, statistical measure technique~\cite{Saad2015}, or employing the machine learning techniques~\cite{Xiwen2019,Grimaldi2019}. A simple interference detection based on \cite{jun2005} is explained in the following. Consider the IQ measurement at the reflector for $k$-th tone (corresponding to frequency $f_k$)\footnote{\R{Note that the same mechanism can be used at the initiator to detect the interfered tones.}}. The reflector downmixes the tone at frequency $f_k$ with its LO (with frequency $f_k$) and takes multiple samples. Then, the samples are averaged, giving $IQ_R(f_k)$. In the absence of interference, the measured envelop is constant for all samples. However, with interference the measured envelop vary over samples. Such variation can be used to detect the interfered tone. This variation can be captured in \cite[eq.~4.1]{jun2005} metric. If \cite[eq.~4.1]{jun2005} metric for a given tone is less than a threshold, then the corresponding IQ is marked as interfered. Furthermore, as IQs of both initiator and reflector impacts the ranging, if the IQ of the either of initiator or reflector is marked as interfered, such tone is considered as interfered.} 
	
	A straightforward approach for range estimation is to zero-pad IQs of the missing/interfered tones. As we will show in section~\ref{Sec:Sim_r}, such zero padding may result in a large ranging error compared to the range estimation without missing/interfered tones. The reason is that the fundamental constant phase progression from tone to tone that was used to make the Hankel matrix in section~\ref{Sec:SysMod} will be violated by zero padding, yielding mixing of the signal with noise sub-space. 
	
	\begin{figure}[t] \centering 
		\includegraphics[scale=0.78]{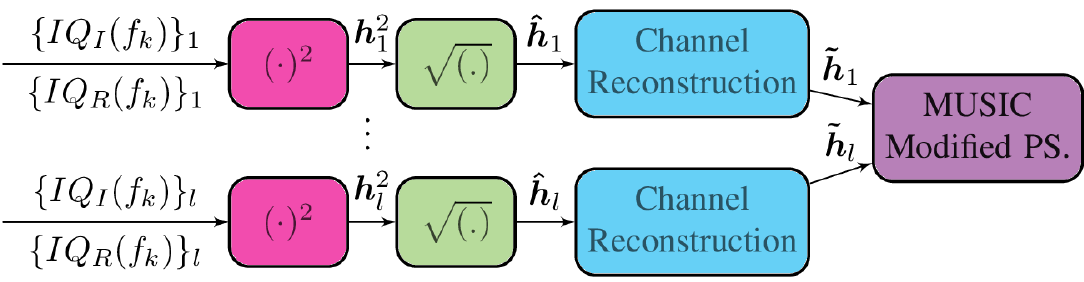}  
		\vspace{-1ex}
		\caption{Schematic of the phase-based ranging with modified \commentJ{Psudo}{pseudo-}spectrum.}  \vspace{-3ex}
		\label{Ranging_Sch1} 
	\end{figure} 
	
	\subsection{Ranging with Modified MUSIC Pseudo Spectrum}\label{Alg1_des}
			
	\R{There are multiple variants of MUSIC in the literature. To reduce the complexity of MUSIC, root-MUSIC \cite{Weiss1993} is proposed which approximate the Eigen value decomposition with polynomial root solver. Gold-MUSIC \cite{Rangarao2013} employs a two stage search to find the peak in the pseudospectrum. The compress-MUSIC \cite{Yan2013} and real-valued MUSIC \cite{Yan2014} partitioned the field-of-view to efficiently estimate the angle-of-arrival and reduce the complexity of the MUSIC. In these papers, it is assumed that the observation (in our case measured IQs at initiator and reflector) are all available. In \cite{Li2017}, an algorithm based on sparsity enforcing norm is proposed to estimate the angle-of-arrival, however, it assumes that the sources are well-separated and the number of sources for angle of arrival is known. This is not a realistic assumption in our problem, as the number of multipath components is a-priori unknown and they can be arbitrary close (not well-separated). To the best of our knowledge, there is no work in literature that modifies the MUSIC pseudospectrum in order to infuse available observations when the full observation is incomplete. Other related works are in essence similar to the method that will be discussed in the next subsection. In the following, we propose two modified cost functions for pseudospectrum which employ the IQ values of the available tone bands to estimate the range\footnote{\R{We highlight that we considered other approaches to infuse the IQ values of the available tone bands for range estimation, e.g., run the MUSIC on each available tone band and then compute the average or weighted average of estimated ranges as the final range estimate. However, they perform worse than the proposed schemes in this subsection in multipath channel. Therefore, to make the paper concise, we excluded them from this subsection.}}.}
	
	The schematic of system model in this subsection is shown in Fig.~\ref{Ranging_Sch1}. As it can be seen, the structure is similar to section~\ref{Sec:SysMod} with the difference that \commentAFF{pseudospectrum of MUSIC is changed}{MUSIC is used with a modified pseudo spectrum}. \commentAFF{}{In what follows, we propose two modified pseudospectrums.} 
	
	Let us assume that there are $l$ available tone bands. For each available tone band the reconstructed channel from the IQ samples of the initiator and reflector can be computed, similar to section~\ref{Sec:SysMod}. We assume that the reconstructed channels of tones in $j$-th available tone band ($j \in \{1,\cdots,l\}$) are denoted as $\boldsymbol{\tilde{h}_j}=[\tilde{h}_{a_j},\tilde{h}_{a_{j+1}},\cdots,\tilde{h}_{b_j}]$. \RRR{$a_j$ and $b_j$ are the first and last index of the $j$-th available tone band, hence, the number of tones in the $j$-th available tone band is ${b_j}-{a_j}+1$.} We define $L_j$ as the smoothing factor of the $j$-th available tone band. For each available tone band the Hankel matrix can be computed as 
	\begin{equation}\label{eq:Hankel_jthband}
		\boldsymbol{H}_j = \left[ {\begin{array}{*{20}{c}}
				{{\tilde{h}_{a_j}}}& \cdots &{{\tilde{h}_{{a_j}+L_j-1}}}\\
				\vdots & \ddots & \vdots \\
				{{\tilde{h}_{{b_j}-L_j+1}}}& \cdots &{{\tilde{h}_{b_j}}}
		\end{array}} \right].
	\end{equation}
	%As explained in section~\ref{Sec:SysMod}, the number of Eigen values are equal to the smoothing factor. 
	We now explain how to select the smoothing factor for the $j$-th available tone band. 
	
	%Consider the $j$-th available tone band. In extreme cases, if the smoothing factor is $1$ or ${b_j}-{a_j}+1$, the number of Eigen values are $1$, hence, signal-noise subspace separation is not possible. Intuitively, it is preferable to have the largest number of Eigen values, as signal-noise subspace separation is easier. In proposition~1, the value of $L_j$ is derived.
	
	\begin{prop} 
		$L_j=\lfloor \frac{{b_j}-{a_j}+1}{2}\rfloor +1$ maximizes the \commentJ{possible}{number of} path delays that \commentJ{can be estimated by finding the}{can be found by} MUSIC \commentJ{Psudo}{Pseudo} spectrum peaks for the $j$-th available tone band.
	\end{prop}\label{prop1}
	
	\begin{IEEEproof} 
		See Appendix~\ref{APP}
	\end{IEEEproof}
	%In Appendix~\ref{APP}, we show that $L_j=\lfloor \frac{{b_j}-{a_j}+1}{2}\rfloor +1$ maximizes the number of Eigen values of ${\boldsymbol{H}_j^H}\boldsymbol{H}_j$, corresponding to the $j$-th available tone band. 
	\R{In Sec.~\ref{Sec:Sim_r}, we verify the Proposition~1 with simulations.}
	
	Now it remains to define \commentAFF{a pseudo spectrum to infuse the signal-noise sub-space separation of different available tone bands. We propose two different pseudo spectrum explained as follows.}{the two modified pseudo spectrum to infuse the signal-noise sub-space separation of different available tone bands.}
	
	\subsubsection{Multiplied pseudo spectrum}	
	We refer to \commentAFF{ranging using system model with multiplied pseudo spectrum as (MPS)}{the first proposed modified pseudo spectrum as multiplied pseudo spectrum (MPS)}. In \commentAFF{this scheme}{MPS}, pseudo spectrum of all available bands will be multiplied to find the peak. Concretely, \commentAFF{}{MPS,} the range is computed by finding the first peak corresponding to the following pseudo-spectrum
	\begin{equation}\label{eq:MPS}
		J(\tau) \triangleq \frac{1}{\prod_{j=1}^l\left({L_j - \boldsymbol{e}{{(\tau )}^H}{{({(\boldsymbol{V}_{S_j})})}^H}({(\boldsymbol{V}_{S_j})})\boldsymbol{e}(\tau )}\right)},
	\end{equation}
	\ASH{where $(\boldsymbol{V}_{S_j})$ is computed by eigenvalue decomposition of ${\boldsymbol{H}_j^H}\boldsymbol{H}_j$}.  Intuitively, finding the first peak in MPS means that all available tone bands should agree that the steering vector of $\tau_0$ ($\boldsymbol{e}(\tau_0)$) is orthogonal to the noise sub-space of all available tone bands. 
	
	\subsubsection{Weighted average pseudo spectrum}
	We refer to \commentAFF{ranging using system model with weighted average pseudo spectrum as (WAPS)}{the second proposed modified pseudo spectrum as weighted average pseudo spectrum (WPS)}. Concretely, the range is computed by finding the first peak corresponding to the following pseudo spectrum
	\begin{equation}\label{eq:WAPS}
		J(\tau) \triangleq \frac{1}{\sum_{j=1}^lL_j\left({L_j - \boldsymbol{e}{{(\tau )}^H}{{({\boldsymbol{V}_{S_j}})}^H}({(\boldsymbol{V}_{S_j})})\boldsymbol{e}(\tau )}\right)}.
	\end{equation}
	In \eqref{eq:WAPS}, the orthogonality of the steering vector of $\tau_0$ ($\boldsymbol{e}(\tau_0)$) are weighted based on the smoothing factors. The motivation for such weighting is that the number of eigenvalues of available tone band with larger smoothing factor selected based on proposition~1 is larger, therefore, signal-noise sub-space separation can be performed more reliably using the thresholds on the eigenvalues.

	\subsection{Ranging with recovering the missing/interfered tones}\label{chan_recov}
	
\begin{figure}[t] \centering 
		\includegraphics[scale=0.58]{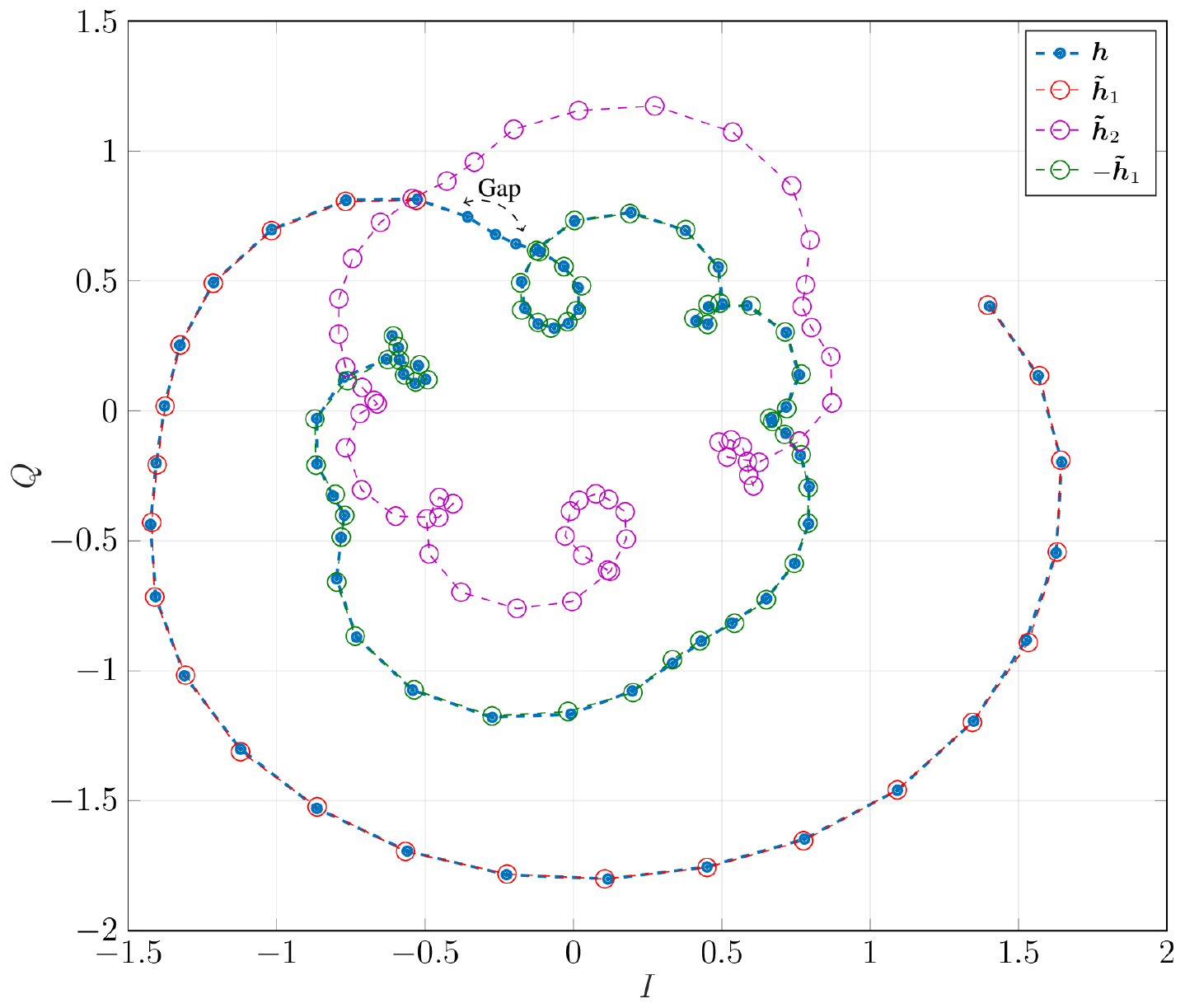}  
		\vspace{-1ex}
		\caption{Comparing the channel reconstruction of two available tone bands.}  \vspace{-1ex}
		\label{hsq_vs_h} 
	\end{figure}

\begin{figure}[t] \centering 
		\includegraphics[scale=0.7]{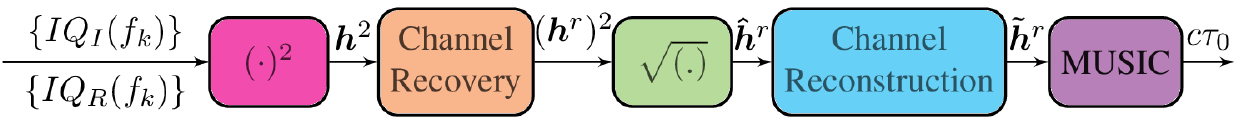}  
		\vspace{-3ex}
		\caption{Schematic of the phase-based ranging with channel recovery.}  \vspace{-4ex}
		\label{Ranging_recov} 
	\end{figure}

	\commentAFF{In this section, we propose two schemes which first estimate the missing/interfered tones and then estimates the range using the MUSIC algorithm. We refer to such estimation as channel recovery in this paper. Recall the system model in Fig.~\ref{Ranging_Super}.}{The schematic of a phase-based ranging with recovering the missing/interfered tones is shown in Fig.~\ref{Ranging_recov}. The schematic in Fig.~\ref{Ranging_recov} is similar to Fig.~\ref{Ranging_Super}, except the channel recovery unit. The channel recovery unit estimates the missing/interfered tones.} For the channel recovery there are two options: (i) recovering of the $\boldsymbol{\tilde{h}}$ corresponding to missing/interfered tones (ii) recovering of the $\boldsymbol{{h}^2}$ corresponding to missing/interfered tones. Option (ii) is preferred over option (i), as there is an ambiguity associated with option (i). The reason of ambiguity for option (i) is that the channel reconstruction in \cite{yao2021ranging} can only \commentJ{grantee}{guarantee} the phase progression within a given available tone band. However, there is no \commentJ{grantee}{guarantee} that the phase progression is correct between two reconstructed available tone bands. To clarify more we provide an example in the following. 
	\begin{Ex}
		Let us assume that the channel jump \commentAFF{}{(or frequency step)} in 80MHz bandwidth is 1MHz, yielding 80 available tones in 2.4GHz ISM band and three tones corresponding to \RRR{2.425GHz, 2.426GHz, 2.427GHz} are missing. In Fig.~\ref{hsq_vs_h} we show IQ values of channel response ($80$ tones) in a multi-path environment based on Saleh-Venezuela channel model \cite{SV_model} ($\boldsymbol{h}$) as well as the result of channel reconstruction for the two available tone bands, i.e., available tone band $[2.41, 2.424]$ GHz ($\boldsymbol{\tilde{h}}_1$) and $[2.428, 2.480]$ GHz ($\boldsymbol{\tilde{h}}_2$). As it can be seen, although $\boldsymbol{\tilde{h}}_1$ is coherent with $\boldsymbol{\tilde{h}}$, there is a $\pi$ phase shift between $\boldsymbol{\tilde{h}}_2$ and $\boldsymbol{\tilde{h}}$. To avoid such phase ambiguity, we consider recovering of $\boldsymbol{{h}^2}$ corresponding to missing/interfered tones. 
	\end{Ex}
	\commentJ{ }{In the following two subsections, we propose two methods to realize channel recovery.}
	
	\subsection{\commentAFF{Recovering the missing channels}{Channel recovery} using atomic norm minimization}\label{Alg2_des}
	
	In the literature, atomic norm as a sparsity enforcing norm is used to estimate the missing samples of a signal comprising the weighted sum of \ASH{complex exponential} when the signal is uniformly sampled \cite{Tang2013}. As discussed in section~\ref{chan_recov}, we aim to recover the $\boldsymbol{h}^2$. Based on \eqref{eq:Chanmodel}, 
	the squared channel response corresponding to $k$-th tone is given as
	\begin{equation}\label{h_sq}
		{({h_k})^2} = \sum\limits_{n = 0}^{M - 1} {\sum\limits_{p = 0}^{M - 1} {{{\tilde a}_n}{{\tilde a}_p}} } \exp ( - jk\Dw ({\tau _n} + {\tau _p})),
	\end{equation}
	where ${{\tilde a}_n} = {a_n}\exp ( - j{\omega _0}{\tau _n}),{{\tilde a}_p} = {a_p}\exp ( - j{\omega _0}{\tau _p})$. As can be seen from \eqref{h_sq}, the squared channel response is also a weighted sum of complex exponentials. Furthermore, the IQs measured on a uniform grid with the step size of $\Df$. Therefore, we can directly apply the atomic norm to estimate the missing/interfered components of $\boldsymbol{h}^2$. In the following, we briefly explain how the recovering of missing channels using atomic norm is formulated.
	
	Based on \eqref{h_sq}, $\boldsymbol{h}^2$ can be computed as
	\begin{equation}
		{\boldsymbol{h}^2} = {[{({h_1})^2},...,{({h_K})^2}]^T} = \sum\limits_{n = 0}^{M - 1} {\sum\limits_{p = 0}^{M - 1} {{{\tilde a}_n}{{\tilde a}_p}} } \boldsymbol{z}({\tau _n} + {\tau _p}),
	\end{equation}
	where $\scalemath{0.79}{\boldsymbol{z}({\tau _n} + {\tau _p}) = {[1,\exp ( - j\Dw ({\tau _n} + {\tau _p})),...,\exp ( - j\Dw K({\tau _n} + {\tau _p}))]^T}}$.
	The set of $\boldsymbol{z}({\tau _n} + {\tau _p})$ for different $\tau _n$ and \commentJ{$\tau _n$}{ $\tau _p$} are known as atoms, which are the building block of ${\boldsymbol{h}^2}$. Let us define $\Omega$ as the set comprising the components of ${\boldsymbol{h}^2}$,\commentJ{ where there is no missing/interfered tones}{ when there are no missing/interfered tones}. Based on Schur Complement and Vandermonde decomposition lemmas \cite{Tang2013},\cite{Chandrasekaran2012}, the missing/interfered tones of $\boldsymbol{h}^2$ can be computed by minimizing the atomic norm of $\boldsymbol{h}^2$ which can be written as a semi-definite programming problem given as 
	\begin{align}\label{SDP}
		\begin{array}{l}
			{\boldsymbol{h}^2} = {\inf _{\boldsymbol{u},t,{\bar{\boldsymbol{h}}^2}}}\left\{ {\frac{1}{{2|{{\bar{\boldsymbol{h}}}^2}|}}trace(Toep(\boldsymbol{u}) + \frac{1}{2}t)} \right\}\\
			\;\;\;\;\;\;\;\;\;\;\;\;\;\;\;\;\; s.t.\left[ {\begin{array}{*{20}{c}}
					{Toep(\boldsymbol{u})}&{{{\bar{\boldsymbol{h}}}^2}}\\
					{{{({{\bar{\boldsymbol{h}}}^2})}^H}}&t
			\end{array}} \right] \succ 0\\
			\;\;\;\;\;\;\;\;\;\;\;\;\;\;\;\;\;\;\;\;\;\;\;\;\;\;\;\;\;\; \bar{\boldsymbol{h}}_\Omega ^2 = \boldsymbol{h}_\Omega ^2,
		\end{array}
	\end{align}
	\ASH{where} $\boldsymbol{u}$, $t$, $\bar{\boldsymbol{h}}^2$ are the optimization variables. The problem of \eqref{SDP} can solved using standard semi definite programming (SDP) solvers such as CVX \cite{cvx}. {\R{We remark that there are other problem formulation for estimating the missing/interfered tones based on de-noising of the atomic norm \cite{Bhaskar2013_denoise} or formulating as a matrix completion problem using nuclear-norm minimization \cite{candes2012exact}. They both boils down to an SDP problem. In the application of range estimation based on phase-based ranging with missing/interfered tones, we did not see any notable performance difference between them, hence, we do not explain these formulations.}}
	
	\subsection{\commentAFF{Recovering the missing channels}{Channel recovery} using Neural Network}\label{Alg3_des}
	
	As\commentJ{it will be seen}{ will be shown} in section~\ref{Sec:Comp_complexity}, the complexity of SDP solver required for the atomic norm minimization is \commentJ{very}{}high, which \commentJ{may prevent}{prevents} its use in a real-time application such as Bluetooth chips with limited \commentJ{hardware}{resources}. Therefore, in this section, we use \commentA{a}{} neural network (NN) to estimate the frequency response of missing/interfered tones which has \commentJ{much}{significant} lower complexity than atomic norm minimization. The idea of \commentA{the}{} NN is to estimate the missing/interfered components of $\boldsymbol{h}^2$ based on the components before and after the tone gap. This is motivated by the fact that there is a progression from components of $\boldsymbol{h}^2$ before the tone gap to the components of $\boldsymbol{h}^2$ after the tone gap. Therefore, we propose to train NN to extract this progression. 
	%Note that as can be seen from Fig.~\ref{Ranging_recov}, the progression of IQ values can have an arbitrary shape which varies over $\boldsymbol{h}$ components due to \comment{multiupath}{multipath} delays. 
	We use feed\commentJ{ }{-}forward NN as it is known to be a universal function approximator \cite{Cybenko1987}, hence, it can estimate such progression.  
	
\begin{figure}[t] \centering 
		\includegraphics[scale=0.7]{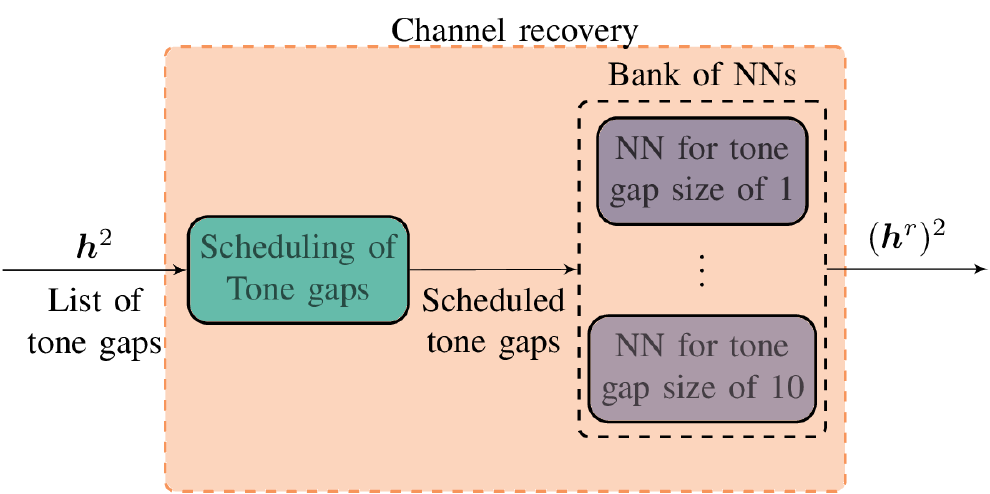}  
		\vspace{-1ex}
		\caption{The schematic of channel recovery block using the NN.}  \vspace{-1ex}
		\label{NN_sys} 
	\end{figure}

	The schematic of channel recovery block using the NN is shown in Fig.~\ref{NN_sys}. As it can be seen, input of the channel recovery block is $\boldsymbol{h}^2$ and a list of tone indices corresponding to tone gaps. First, the list of gaps are ordered based on a scheduling algorithm, which will be explained in section~\ref{sec:schedule}. Then, corresponding to number of missing/interfered tones in a tone gap, a trained NN out of a bank of NNs is selected \commentJ{which estimates}{to estimate} the \commentJ{squared}{two-way} frequency response of missing/interfered tones. After recovering \commentJ{the all tone}{all} gaps, $(\boldsymbol{h}^{r})^2$ is build which will be passed to channel \commentJ{reconstrunction}{reconstruction} and MUSIC algorithm as shown in Fig~\ref{Ranging_recov}. In the following, we first explain the structure of the neural network. Then, we discuss the scheduling algorithm.
	
	\subsubsection{Structure of each NN in the bank of NNs}
	
	In the following, we explain the procedure for training of \RRR{a NN}, where the number of missing/interfered tones in a tone gap is $W$. For the input features of NN, we use some components of $\boldsymbol{h}^2$ corresponding to tones before and after the tone gap. The number of such components of $\boldsymbol{h}^2$  is a design parameter, which provides a performance-complexity trade-off of using the NN. Empirically, we found that it is enough to use $W$ components of $\boldsymbol{h}^2$ before and after the tone gap. \commentA{}{\it Feels logical that there is a relation with the coherence bandwidth!} We consider a real-valued NN, meaning that the weights, input, and output are real numbers. For such NN, the gap of width $W$ results in $4W$ input and $2W$ output neurons. \commentA{$2W$/$2W$}{\it I dont understand this notation?} inputs correspond to the real and imaginary parts of $W$ components of $\boldsymbol{h}^2$ \ASH{before/after} the tone gap. %Similarly,  $2W$ inputs correspond to the real and imaginary parts of $W$ components of $\boldsymbol{h}^2$ after the tone gap. 
	Furthermore, $2W$ outputs correspond to the components of  $\boldsymbol{h}^2$ at the tone gap. To further limit the complexity, we only consider one hidden layer with $20$ neurons. Note that this number is empirically found for the tone gap  with $10$ missing/interfered tones. It is also worth \commentJ{to point out}{noting} that for the considered NNs (NNs for the tone gaps with number of missing tones up to $10$), $20$ is the minimum number of neurons for the hidden layer that prevents the bottleneck \cite{Bai2015AnalysisOA}. \RR{The parameters regarding the NN structure and training are summarized in Table~\ref{Tabcomp1}.} The input and output of NN are normalized properly such that the mean of the input and output layer is zero and the corresponding variance is $1$. This normalization is known to improve the convergence speed of the NN training \cite{Ioffe2015}, \cite{Huang2020NormalizationTI}. 
	
%	\begin{table*}[t]
%		\caption{Codes used for simulations}
%		\centering
%		\begin{tabular}{ccccccccccccc}
%			\makecell{Component\\ code} &
%			\makecell{component\\ parameters} &
%			\makecell{component\\ code rate} &
%			\makecell{PC \\ code rate} &
%			\makecell{HD Shannon limit  \\ at PC code rate} & 
%			\makecell{SD Shannon limit \\ at PC code rate} &
%			\makecell{SCC \\ code rate} &
%			\makecell{HD Shannon limit \\ at SCC code rate} &
%			\makecell{SD Shannon limit \\ at SCC code rate} \\
%			\midrule
%			$\mathcal{C}_1$ & (256,239,2) & 0.933 & 0.871 & 4.05 (dB) & 2.64 (dB) & 0.867 & 3.99 (dB) & 2.74 (dB) \\
%			$\mathcal{C}_2$ & (255,231,3) & 0.905 & 0.820 & 3.54 (dB) & 2.23 (dB) & 0.811 & 3.46 (dB) & 2.14 (dB)\\
%			$\mathcal{C}_3$ & (511,484,3) & 0.947 & 0.897 & 4.36 (dB) & 3.15 (dB) & 0.894 & 4.32 (dB) & 3.11 (dB) \\
%			\midrule
%		\end{tabular}
%		\label{codes}
%	\end{table*}

\begin{table}[t]	
	\caption{\RR{The parameters of NN structure and the corresponding training}}
	%\vspace{-2ex}
	\centering
	\renewcommand{\arraystretch}{1.1}
	\scalebox{1.1}{	
		\begin{tabular}{c|c}
			\toprule
			\midrule
			\RR{Activation functions:} & \makecell{\RR{ReLU for hidden layer,}  \\ \RR{no activation for the output layer}}  \\
			\RR{Cost function:} & \RR{minimum mean squared error}\\ 
			\RR{Optimizer method:} & \RR{ADAM optimizer \cite{Diederik_DBLP}}\\ 		
			\RR{Learning rate:} & \RR{0.001} \\ 		
			\RR{Batch size:} & \RR{1000} \\
			\RR{number of epochs:} & \RR{50}\\ 								
			\midrule
			\bottomrule
		\end{tabular}
	}
	%\vspace{-0.4cm}
	\label{Tabcomp1}
	%\vspace{-2ex}
\end{table} 
	
	%For the activation functions we use the “\commentJ{Relu}{ReLu}” function for the hidden layer. Furthermore, as the output layer gives the components of $\boldsymbol{h}^2$ corresponding to tone gap, no activation function is used for the output layer. 
	
	%For training\commentJ{ of each NN}{}, we use the \commentJ{minimum mean squared cost function}{mean squared error as cost function}. We also use the ADAM optimizer \cite{Diederik_DBLP} with a learning rate of $0.001$. Furthermore, a batch size of $1000$ training samples and $50$ epochs are used during the training phase. 

	%Let us assume that $\boldsymbol{x}$ is an input training example of length $4W$. Furthermore, $\boldsymbol{y}$ is an output training example of length $2W$. We denote by $\boldsymbol{\mu}_\text{in}$ and $\boldsymbol{\sigma}_\text{in}$ are vectors of length 4W corresponding to mean and standard deviation of all input training examples. Furthermore, assume that  $\boldsymbol{\mu}_\text{out}$ and $\boldsymbol{\sigma}_\text{out}$ are vectors of length 2W corresponding to mean and standard deviation of all output training examples. The normalized input and output training examples are denoted by $\boldsymbol{x}_\text{normalized}$ and $\boldsymbol{y}_\text{normalized}$, respectively, which are given as  
	%\begin{align}
	%& \nonumber \boldsymbol{x}_\text{normalized} = diag(\boldsymbol{\sigma}_\text{in})^{-1}(\boldsymbol{x}-\boldsymbol{\mu}_\text{in}) \\ 
	%& \boldsymbol{y}_\text{normalized} = diag(\boldsymbol{\sigma}_\text{out})^{-1}(\boldsymbol{y}-\boldsymbol{\mu}_\text{out})
	%\end{align}
	
\begin{figure}[t] \centering 
		\includegraphics[scale=1.75]{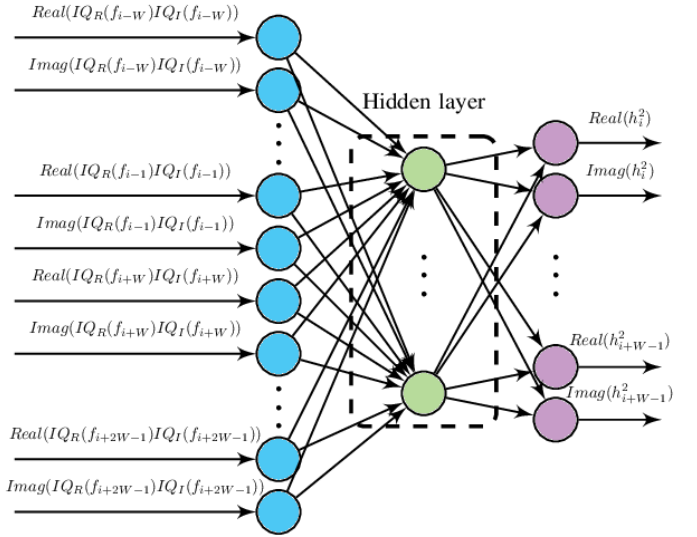}  
		\vspace{-2ex}
		\caption{The schematic of NN for recovering  $[{({h_i})^2},{({h_{i+1}})^2},...,{({h_{i+W-1}})^2}]^T$.}  \vspace{-1ex}
		\label{NN_struc} 
	\end{figure}

	\commentJ{To train the NN}{To train the NN}, we generate gaps of width $W$ with a random position \commentJ{on}{in} the spectrum. Then, for each gap realization, we generate the frequency channel response and IQ samples based on the SV channel model~\cite{SV_model}. Then, an exponent with \commentJ{uniformly}{uniformly distributed} random phase (see $\theta_k$ in \eqref{eq:IQ}) is added to IQs to mimic the incoherent frequency switching. \RRR{We remark that due to noise at initiator and reflector the noise is also added to the IQ samples (given in \eqref{eq:IQ}). Therefore, $h_k^2 = (I{Q_R}({f_k})+n_R(({f_k})))(I{Q_I}({f_k})+n_I(({f_k})))$, where $n_R(({f_k}))$ and $n_I(({f_k}))$ stand for the noise realization at the $k$-th tone for initiator and reflector, respectively.} The parameters of SV model (which will be discussed in Section~\ref{Sec:Sim_r}) are also randomly selected to mimic different fading conditions. This random selection of the parameters of simulation model is essential in the training stage, as the trained model is expected to operate in any unseen ranging setup and fading conditions. More details on parameters used for generating the training data set are explained in Section~\ref{Sec:Sim_r}. The schematic of NN for recovering  $[{({h_i})^2},{({h_{i+1}})^2},...,{({h_{i+W-1}})^2}]^T$ is shown in Fig.~\ref{NN_struc}.
	
	\subsubsection{Scheduling of the tone gaps}\label{sec:schedule}
	
	\begin{algorithm}[t]\label{Alg1}
		\small
		\SetKwInOut{Input}{Input}\SetKwInOut{Output}{Output}\SetKwInOut{Ini}{Initialization}\SetKwInOut{Inter}{Variables}
		\SetAlgoLined
		\Input{\emph{input~tone~gap~map}, \emph{available~tone~indices}}
		\Output{\emph{output gap map}}
		\Inter{\emph{current gap map, scheduled~gaps, not~scheduled~gaps, old~output~gap~map}}
		\Ini{\emph{current~gap~map} $\leftarrow$ \emph{input~tone~gap~map} \\ \emph{output~gap~map} $\leftarrow$ $\{\}$ \\ \emph{scheduled~gaps} $\leftarrow$ $\{\}$ \\ \emph{not~scheduled~gaps} $\leftarrow$ $\{\}$ \\ \emph{old~output~gap~map} $\leftarrow$ $\{\}$}
		\While{current~gap~map \text{\upshape is not empty}}{
			\For{$j$ $\in$ current~gap~map}{
				\eIf{\text{\upshape the required inputs for tone gap $j$ is in} available~tone~indices}{
					\text{Add the tone gap $j$ to} \emph{scheduled~gaps}\;
				}{
					\text{Add the tone gap $j$ to} \emph{not~scheduled~gaps};
				}
			}
			\uIf{not scheduled gap \text{\upshape is empty}}{
				\begin{itemize}
					\item Add tone gaps in \emph{scheduled gaps} to \emph{output gap map}
					\item \emph{current~gap~map} $\leftarrow$ $\{\}$
				\end{itemize}
			}
			\uElseIf{not scheduled gaps $==$ old not scheduled gaps}{
				\begin{itemize}
					\item \text{\upshape Sort the gaps in} \emph{not scheduled gaps}  \text{\upshape based on their number of tones in an increasing order} 
					\item \text{\upshape Add sorted gaps to} \emph{output gap map}
					\item \emph{current~gap~map} $\leftarrow$ $\{\}$
				\end{itemize}
			}
			\Else{
				\begin{itemize}
					\item \emph{current~gap~map} $\leftarrow$  \emph{not~scheduled~gaps}
					\item \text{\upshape Add gaps in} \emph{scheduled~gaps} \text{\upshape to the} \emph{output~gap~map}
					\item \text{\upshape Update} \emph{available~tone~indices} \text{\upshape by adding indices of} \\ \emph{scheduled~gaps}
					\item  \emph{old~not~scheduled~gaps} $\leftarrow$ \emph{not~scheduled~gaps}
				\end{itemize}
			}
			
		}
		\caption{Scheduling of tone gaps}
	\end{algorithm} 
	
	Let us assume that different NNs, each corresponding to a given number of missing/interfered tones are trained, forming a bank of NNs. For a given tone gap, performance of the channel recovery based on NN depends on which order the components of the bank of NNs are used. %This is due to the fact that the input features of each NN may depend on the recovered components of $\boldsymbol{h}^2$ corresponding to the output of a NN for another gap. 
	To clarify the impact of order in recovering of the tone gaps, in the following, we provide an example. 
	
	\begin{Ex}
		For the ease of explanation, we refer to components of $\boldsymbol{h}^2$ with the corresponding indices. For instance, for channel jump of $1$MHz in 80MHz bandwidth in 2.4GHz ISM band, there are $80$ tones from frequency $2.401$GHz to $2.48$GHz, which we referred with indices $\{0,\cdots,79\}$. Let us assume that there are four tone gaps with indices $\{24, 25, 26\}$, $\{29,30\}$, $\{32\}$, and $\{34, 35\}$. The NN for recovering the components of $\boldsymbol{h}^2$ for $\{24, 25, 26\}$ requires the IQ samples at the index $29$, which is not available due to tone gap $\{29,30\}$.  one can easily infer that it is better to recover tone gap $\{29,30\}$ earlier than tone gap $\{24, 25, 26\}$, as the recovered value of $\boldsymbol{h}^2$ at index $29$ can be used for recovering of $\{24, 25, 26\}$. With the same reasoning, one can check that it is better to recover $\{32\}$ earlier than $\{29,30\}$ and $\{34, 35\}$. Following the same reasoning, the order for recovering the gaps is found as $\{32\}$, $\{29,30\}$, $\{34, 35\}$, and $\{24, 25, 26\}$.
	\end{Ex} We propose an scheduling algorithm in Algorithm~1, which provides the order of recovering of tone gaps. The algorithm is based of the following two heuristics:
	\begin{itemize}
		\item It is preferred to first recover the tone gaps, where the corresponding inputs of NN are not missing/interfered. The reason is that recovering of such gaps may provide some inputs for NNs that will recover other tone gaps.
		\item In the case that recovering \commentJ{each}{} of two tone gaps \commentJ{are dependent to each other}{rely on each other} (e.g., recovering of tone gaps with indices $\{17, 18\}$ and $\{21, 22, 23, 24\}$), it is better to first recover the channel gap with smaller size by zero padding the values of $\boldsymbol{h}^2$ corresponding to the larger tone gap. The reason is that the performance of channel recovery using NN degrades \commentJ{by increasing the}{with increasing} number of missing/interfered tones. 
	\end{itemize}
	
	The input of Algorithm~1 is the “input tone gap map”, which is a list containing the indices of tone gaps. The output of the algorithm is “output gap map” which contains the scheduled tone gaps. Furthermore, Algorithm~1 requires “available tone indices” \commentJ{is a list for}{as a list of} indices of tones which are not missing/interfered. For the example given in this section, the input tone gap map is $\{\{24, 25, 26\}, \{29, 30\}, \{32\}, \{34, 35\}\}$ and available tone indices are $\{0:23, 27, 28, 31, 33, 36:79\}$.    
	
	\section{\RR{Computational Complexity of the Channel Recovery Schemes}}\label{Sec:Comp_complexity}
			\begin{table*}[t]	
		\caption{\R{Comparing the computational complexity and required flash memory of the proposed schemes for ranging with missing/interfered tones w.r.t. based line  when there is no missing/interfered tone}}
		%\vspace{-2ex}
		\centering
		\renewcommand{\arraystretch}{0.5}
		\scalebox{0.9}{	
			\begin{tabular}{c|cc}
				\toprule
				\R{Method} & \R{\makecell{Computational complexity \\ (flops) }}  & \R{\makecell{Extra required memory w.r.t. base line \\ (number of floating point values)} } \\	
				\hline \\
				\R{\makecell{Base line: MUSIC without\\ missing/interfered tones} } & \R{$\mathcal{O}((\lfloor \frac{K}{2}\rfloor +1)^3)$} & \R{$\text{N.A}$} \\	
				\hline \\
				\R{\makecell{MUSIC with \\ modified pseudospectrum}} & \R{$\sum\limits_{i = 1}^{l} {\mathcal{O}((L_j)^3)}$} & \R{None} \\	
				\hline \\						
				\R{\makecell{MUSIC with channel recovery \\ using atomic norm minimization \eqref{SDP}  \cite{Tang2013} \\ or formulation \cite{Chandrasekaran2012} or nuclear norm \cite{candes2012exact}}} & \R{\makecell{$\mathcal{O}((\lfloor \frac{K}{2}\rfloor +1)^3)+\mathcal{O}(K^6)$ (IMP opt. \cite{HANSEN20197}) \\ $\mathcal{O}((\lfloor \frac{K}{2}\rfloor +1)^3)+100 \times \mathcal{O}(K^3)$ (ADMM opt. \cite{Bhaskar2012ex})}} & \R{None} \\
				\hline \\		
				\R{\makecell{MUSIC with channel recovery \\ using NN}} & \R{$\mathcal{O}((\lfloor \frac{K}{2}\rfloor +1)^3) + (12WH+12W)$} & \R{$\sum\limits_{i = 1}^{T} {(6Hi+(H+2i)+12i)}$}  \\															
				\bottomrule
			\end{tabular}
		}
		%\vspace{-0.4cm}
		\label{comp_complex}
		\vspace{-4ex}
	\end{table*} 

	In this section, we evaluate the additional computational complexity of the proposed schemes in \RRR{Sections~\ref{Sec:ranging_mis_interf}} compared \commentJ{with the system, where there is no missing/interfered tones }{to a system with no channel recovery scheme (see \ref{MUSIC_general})}. We evaluate the complexity based on floating\commentJ{ }{-}point operations (flops) \cite{HANSEN20197}. We highlight that the computational complexity of ranging without missing/interfered tones (see Fig.~\ref{Ranging_Super}) is mainly dominated by the complexity of the eigenvalue decomposition in the MUSIC algorithm. The complexity of eigenvalue decomposition in \eqref{eq:SN} using proposition~1 is $\mathcal{O}((\lfloor \frac{K}{2}\rfloor +1)^3)$. 
	\subsection{\RR{Ranging with modified MUSIC pseudo spectrum}}
	In the scheme proposed in Section~\ref{Alg1_des}, only eigenvalue decomposition for the available \commentJ{ton bands}{sub-bands} are required, hence, the complexity of system is sum of complexity of eigenvalue decomposition on each available tone band. We remark that the complexity of eigenvalue decomposition for $j$-th band with smoothing factor of $L_j$ is $\mathcal{O}((L_j)^3)$. As $L_j<\lfloor \frac{K}{2}\rfloor +1$, the scheme proposed in Section~\ref{Alg1_des}, does not impose an extra complexity compared to the system, \RRR{where there are} no missing/interfered tones. \R{In this scheme, no extra memory w.r.t. system with no channel recovery scheme (see \ref{MUSIC_general}) is required.}  
	\subsection{\RR{Channel recovery based on using atomic norm minimization}}
	\R{In the scheme proposed in Section~\ref{Alg2_des} \R{(based on atomic norm with or without de-noising \cite{Tang2013},\cite{Chandrasekaran2012} or nuclear norm \cite{candes2012exact}) minimization}, the channel recovery is based on solving an SDP problem.} The SDP solvers typically solve \eqref{SDP} using the interior point method (IPM) \cite{cvx}. The complexity of the IPM is at least $ \mathcal{O}(K^6)$ flops per iteration at best \cite{HANSEN20197}. To reduce the complexity of IPM, different solver known as alternating direction method of multipliers (ADMM) can be employed which is known to have less accuracy than IPM \cite{Boyd2011}. The complexity of ADMM per each iteration is dominated by the eigenvalue decomposition of a $K \times K$ matrix, which requires $\mathcal{O}(K^3)$ flops \R{\cite{Bhaskar2012ex}}. Note that the number of iterations required for SDP solvers to converge to the solution is around $100$. Comparing $100 \times \mathcal{O}(K^3)$ with $ \mathcal{O}((\lfloor \frac{K}{2}\rfloor +1)^3)$, one can conclude that the additional complexity of ranging based on channel recovery in Section~\ref{Alg2_des} is much larger than the complexity of ranging without missing/interfered tones. \R{In this scheme, no extra memory w.r.t. system with no channel recovery scheme (see \ref{MUSIC_general}) is required.}  
	
	\subsection{\RR{Channel recovery using neural network}}
	The complexity of the scheme proposed in Section~\ref{Alg3_des} is twofold: (i) arithmetic operations required for forward propagating through components of the bank of NNs, (ii) the memory required to store the parameters of the bank of NNs. We consider that the bank of NNs contains $T$ different NNs, where the first NN is trained for recovering tone gap with number of missing/interfered tones of $1$ and the last NN is trained for recovering tone gap with number of missing/interfered tones of $T$. Furthermore, one hidden layer with $H$ neurons. 
	
	For forward propagating in the NN trained for recovering the gap number of missing/interfered tones of $W$, the required number of real multiplications and real additions required are $6WH+6W$ and $6WH+6W$, respectively. $6WH+6W$ is computed as follows: $4WH$ real multiplications/additions are required for propagating from the input layer to the hidden layer, $2WH$ real multiplications/additions are required for propagating from the hidden layer to the output layer, $4W$ real multiplications/additions are required for the input layer normalization, and $2W$ real multiplications/additions are required for the output layer normalization. \R{In total, the scheme requires $12WH+12W$ flops}. Furthermore, the total number of real values that should be stored as the parameters of the components of bank of NNs is 
	\begin{equation}\label{comp_NN}
		\sum\limits_{i = 1}^{T} {(6Hi+(H+2i)+12i)}
	\end{equation}
	In \eqref{comp_NN}, $6Hi$ is the total number of stored weights required for forward propagating from input to output of the NN trained for tone gap with number of components of $i$. $H+2i$ is total number of biases in NN trained for tone gap with number of components of $i$.  Furthermore, $12i$ is the total number of values required to be stored for input and output normalization of the NN trained. \R{In Table.~\ref{comp_complex}, we compare the extra computational and memory requirement of proposed scheme in this paper with the baseline when there is no missing/interfered tone and MUSIC is used as a super resolution algorithm for range estimation. In the following example, we provide an example for the complexity comparison.}
	
	\begin{Ex}
		Let us assume $K=80$, $T=10$, $H=20$, and $100$ 
		\RRR{iterations} for ADMM algorithm as a SDP solver, as considered in Section~\ref{Alg3_des}. \R{The complexity of ranging system without missing/interfered tones is $\mathcal{O}(68921)$ (base line in Table~\ref{comp_complex}). The extra complexity of using channel recovery based on atomic norm w.r.t. base line system is $100 \times \mathcal{O}(512000)$, and the extra complexity of using the largest component in the bank of NNs w.r.t. base line system is $2640$. One can see that the extra complexity by NN is negligible.} Furthermore, the total memory required for implementing the bank of NNs for single-precision and double-precision floating-point are $30.2$ Kbytes and $60.5$ Kbytes, respectively. \RR{This required memory is sufficiently small for IoT devices such as BLE.}
	\end{Ex}

	\section{Simulation and measurement Results}\label{Sec:Sim}
	In this section, we evaluate the performance of ranging using schemes in Section~\ref{Sec:ranging_mis_interf}, when there are missing/interfered tones. For the sake of comparison, we also show the performance of ranging \commentJ{when there is}{with} no missing/interfered tones. We also show the performance of ranging system based on Fig~\ref{Ranging_Super}, when the missing/interfered tones are zero padded. We first show the simulation results, then, we show the results based on the measured data. 
	
	\subsection{Simulation results}\label{Sec:Sim_r}
	
	\commentJ{For simulation and also training the NN}{Both for simulation and training of the NN}, we used the SV model \cite{SV_model}. In the following, we briefly explain the model as well as the parameters used for training. The SV models the time difference between two consecutive rays as \commentJ{exponential distribution}{a exponentially distributed random variable} with parameter $\lambda$. \commentA{The mean of such exponential distribution is $\frac{1}{\lambda}$, hence, $\frac{1}{\lambda}$ shows the expected time difference between two consecutive rays.}{} The first time delay ($\tau_0$) is generated based on the distance between initiator and reflector. Furthermore, there is a Rician factor parameter, which shows the ratio of power of the LoS component w.r.t. all none LoS components \cite{SV_model}. Each realization of the SV model gives $\{a_m\}$ and $\{\tau_m\}$ in \eqref{eq:Chanmodel}. The IQ values are generated based on a uniform random phase (see \eqref{eq:IQ}). The additive white Gaussian noise is added to the IQ samples in order to have a given signal-to-noise ratio (SNR), which is defined as
	\begin{equation}\label{SNR}
		SNR = \frac{\frac{1}{M}\sum\limits_{n = 0}^{M - 1} {|a_m|^2}}{N_0},
	\end{equation}
	where $N_0$ is the variance of complex Gaussian distributed noise. For training the NN, we generate the training and validation sets based on SV model, where the parameters of the model are uniformly random selected from the following intervals: $SNR \in [20, 30]$ dB, $\tau_0 \in [1, 30]$ ns, $\frac{1}{\lambda} \in [4, 10]$ ns, and Rician factor $\in [-15, 15]$ dB. The \commentA{root mean square of the delay spread of the channel is chosen as $22$ ns. This means that the coherence bandwidth is $7.2$ MHz, therefore, the channel is frequency selective for the available bandwidth $80$ MHz in the ISM frequency band. The frequency selective property of the channel in ISM band is confirmed by measurements \cite{boer2019}}{\it The RMS delay spread missing in both description and value range. Important for the coherence bandwidth of the channel}. The size of training set is $900000$ channel realizations and the size of validation set is $100000$ channel realizations.  
	
	\R{To verify the proposition~1, in Table~\ref{Tab_smooth}, we compare the median of the estimated range (as a measure of the accuracy) using phase-based ranging with MUSIC (Fig.~\ref{Ranging_Super}) for different values of spatial smoothing parameter $L$. In this table, we assume that there are no gaps (all $80$ tones are used) and $L=\lfloor F\times80\rfloor +1$. Furthermore, the Rician factor and SNR are in SV model are considered as $0$ dB and $20$dB, respectively. As it can be seen, the minimum median (best accuracy) is for value of $F=0.5$, which affirms the proposition~1.}

	\begin{table}[t]	
		\caption{\R{Comparing the median of the range estimate for different values of spacial smoothing parameter}}
		%\vspace{-2ex}
		\centering
		\renewcommand{\arraystretch}{0.45}
		\scalebox{0.85}{	
			\begin{tabular}{c|ccccccccc}
				\toprule
				\R{$F$} & \R{$0.1$} & \R{$0.2$} & \R{$0.3$} & \R{$0.4$} &\R{$\bf 0.5$} & \R{$0.6$} & \R{$0.7$} & \R{$0.8$} & \R{$0.9$} \\	
				\hline \\
				\R{$L$} & \R{$9$} & \R{$17$} & \R{$25$} & \R{$33$} &\R{$\bf 41$} & \R{$49$} & \R{$57$} & \R{$65$} & \R{$73$} \\	
				\hline \\
				\R{\makecell{Median of the\\ range estimate [cm]}} & \R{$96$} & \R{$41$} & \R{$22$} &\R{$14$} & \R{$ \bf 10$} & \R{$11$} & \R{$14$} & \R{$15$} & \R{$22$} \\							
				\bottomrule
			\end{tabular}
		}
		%\vspace{-0.4cm}
		\label{Tab_smooth}
		\vspace{-1ex}
	\end{table} 

	\begin{figure}[t] \centering 
	\includegraphics[scale=0.58]{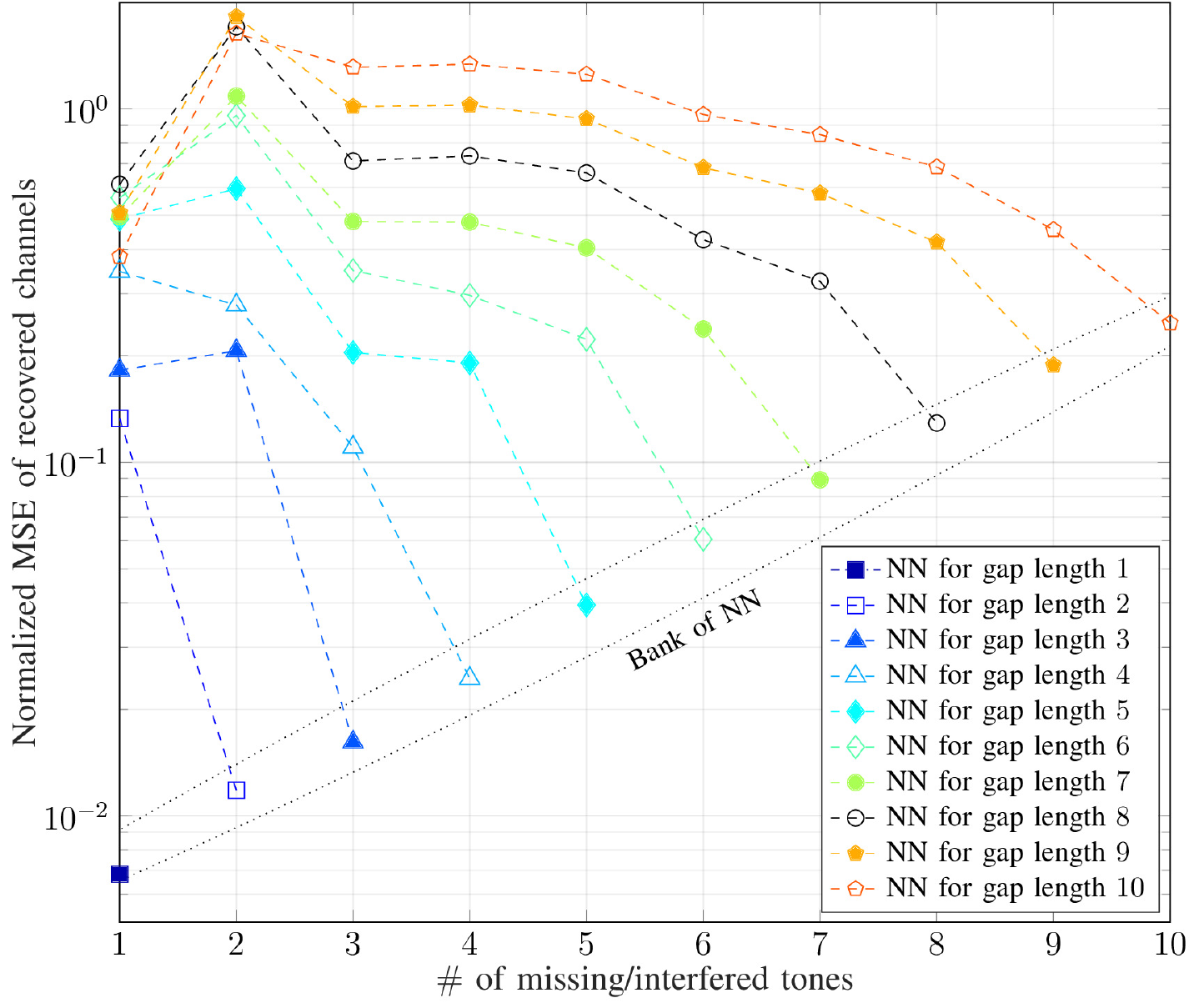}  
	\vspace{-1ex}
	\caption{\R{The normalized MSE of the recovered channels when NN trained for a given gap length is used for gaps of smaller length (ablation study).}}  \vspace{-4ex}
	\label{ablation} 
\end{figure} 

	\begin{figure*}[!htb] \centering 
	\includegraphics[scale=0.7]{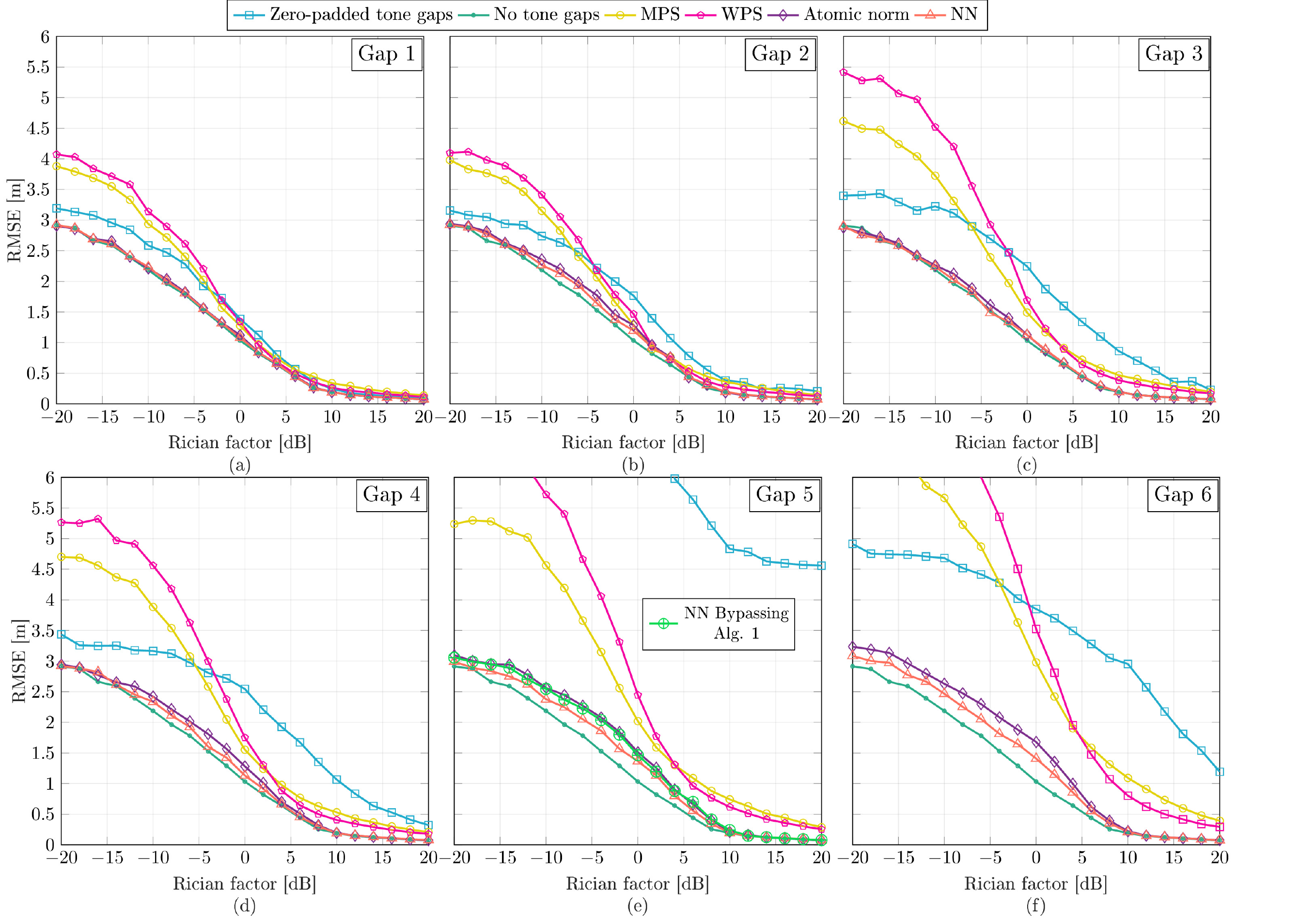}  
	\vspace{-3ex}
	\caption{The RMSE of range estimation using different schemes for (a) gap 1, (b) gap 2, (c) gap 3, (d) gap 4, (e) gap 5, (f) gap 6}  \vspace{-3ex}
	\label{Fig_RMSE} 
\end{figure*}

\R{As an ablation study, in Fig.~\ref{ablation} we show the normalized mean squared error (MSE) of the recovered channels ($(\boldsymbol{h}^{r})^2$ in Fig.~\ref{Ranging_recov}) when NN trained for a given gap length is used to recover the gaps of smaller lengths. Note that if the NN is trained for gap of length $W$, the size of its input and output layer is larger than the size of its input and output layer of a NN trained for gaps of length smaller than $W$ (see Sec. III.D). To be able to compare the MSE for different gap lengths, we normalize the MSE by the gap length. In this figure, for a given gap length, we randomly select the position of the gap in the $80$ tones. As it can be seen, the normalized MSE is minimized when the NN trained for given gap length is used. This verifies the superiority of our proposed scheme in using a bank of NNs for recovering the missing/interfered tones.}
	
	\commentJ{For evaluation of }{To evaluate} the performance of \commentJ{different}{all proposed} schemes, \RRR{for the simulation results} we independently generate channel realizations, where the distance between initiator and reflector is $6$m, $\frac{1}{\lambda}=4$ns, and $SNR=20$dB. We consider channel jump of $1$MHz in $80$MHz bandwidth in $2.4$GHz ISM band, there are $80$ tones from frequency $2.401$GHz to $2.48$GHz, which we referred with indices $\{0,\cdots,79\}$. We consider $6$ different gaps. \RRR{In all of these $6$ gaps, we assume tones with indices $\{0:2\}, \{78:79\}, \{24:26\}$ are the missing tones, therefore, in each tone gap, we consider at least three gaps due to the missing tones. Furthermore, we assume that the indices corresponding to interfered tones of gaps 1-6 are \RRR{$\{\}$, $\{27:29\}$, $\{59:61\}$, $\{59:67\}$, $\{29:30, \; 32, \; 34:36, \; 38, \; 52:55, \; 58:59, \; 62:63, \; 65\}$, and $\{12:13, \; 37:39, \; 40:41, \; 53:55, \; 61:64, \; 70:72\}$}, respectively. Note that the overall number of tone gaps for each tone gap can be computed based on the number of the interfered gaps and missing tones, e.g., the number of gaps for gap 5 is $(8+3=11)$.} \commentJ{Note that the simulation parameters and the considered gaps are just selected for the sake of performance evaluation \ASH{i.e., the conclusions that will be drawn from the simulation results can be also drawn from other simulation parameters and different tone gaps} }{These values are chosen such that we can study the impact of the gap size and its position on the overall ranging performance.} 
	
	%\comment{drawn from the simulation results can be drawn from other simulation parameters/gaps}{\it strange sentence, but I dont know how to fix it, as I dont fully understand what you want to say.}. 

\begin{figure}[!htb] \centering 
	\includegraphics[scale=0.33]{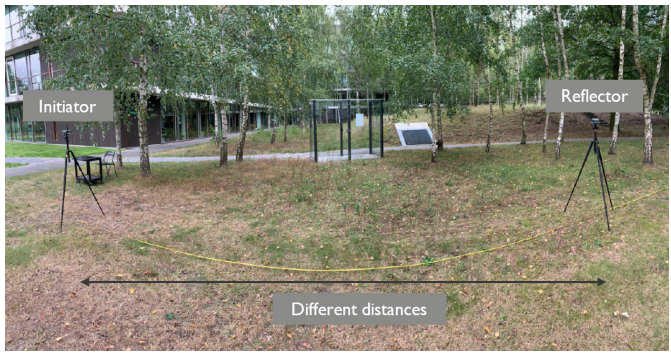}  
	\vspace{-2ex}
	\caption{\R{The wireless outdoor measurement setup using NXP KW36 chip for comparing the range estimation with phase-based ranging using MUSIC (Fig.~\ref{Ranging_Super}) and ToF.}}  \vspace{-4ex}
	\label{setup_MUSIC_vs_ToF} 
\end{figure}

	In Fig~\ref{Fig_RMSE}\commentJ{ we show}{,} the root mean square error (RMSE) of range \commentA{}{General comment about RMSE (more valid for the measurements): if the bias is not removed, and we compare the RMSE, the RMSE is a way of comparing accuracy. Besides, the Q90 is a measure of precision. If we assume that the bias is constant for all the measurement scenarios, my suggestion is to remove the bias, and only compare the Q90, as the accuracy would be the same after bias removal.} estimation error \commentJ{ }{is shown} for gaps $1$-$6$ when the Rician factor varies. As it is expected, by reducing the Rician factor, the RMSE increases. Comparing Fig~\ref{Fig_RMSE}(a) with Fig~\ref{Fig_RMSE}(b), and Fig~\ref{Fig_RMSE}(c) with Fig~\ref{Fig_RMSE}(d), one can see that \commentJ{the by}{with} increasing number of missing/interfered tones, the RMSE of all schemes \commentJ{increase}{increases as well}. However, such increase is less than $8$ cm for schemes based on channel recovery using atomic norm and NN. 
	%Comparing Fig~\ref{Fig_RMSE}(a) with Fig~\ref{Fig_RMSE}(c), with the same number of missing/interfered tones, the RMSE of ranging error increases for the tone gap which is more close to the middle of the $80$MHz bandwidth. This is expected as more components of the Hankel matrix \eqref{label}-\eqref{label} are changed compared to the case where there is no tone gap. Therefore, the quality of the signal-noise subspace separation is lower, i.e., there maybe sum noise subspaces with large Eigen values, which will be considered as signal space. 
	In Fig~\ref{Fig_RMSE}(e) and Fig~\ref{Fig_RMSE}(f), the number of tone gaps \commentJ{increased}{increases from $4$ on Fig~\ref{Fig_RMSE}(b)-(d) to $11$ in Fig~\ref{Fig_RMSE}(e) and $9$ in Fig~\ref{Fig_RMSE}(f)}, yielding less available tone bands. Therefore, the RMSE of the MPS and WPS schemes which depends on the length of available tone bands \commentJ{are also increased}{increases as well}. Furthermore, for tone gap 5, one can \commentJ{see}{observe} that the ranging using zero-padded tone gaps results in \ASH{larger error than other schemes}. This is due to the fact \commentJ{}{that} large number of zeros-padded tone gaps completely changes the signal-noise sub-space separation of the MUSIC algorithm, leading to wrong first path peak in the pseudo spectrum. Furthermore, in Fig~\ref{Fig_RMSE}(e) we show RMSE of the ranging using channel recovery based on NN, when the scheduling algorithm is bypassed. As it can be seen, without scheduling algorithm the RMSE of ranging is increased by up to $35$cm. In particular, it performs roughly the same as ranging with channel recovery using atomic norm minimization. This highlights the impact of the scheduling algorithm on the performance of ranging with channel recovery using the NN.   
	
	Overall, from Fig.~\ref{Fig_RMSE} one can conclude that ranging with channel recovery using NN performs the best compared to other schemes. In particular, if the number of tone gaps and the number of tones in each gap are small (e.g., consider tone gap 1 or tone gap 3), the ranging performance of channel recovery using NN is the same as ranging without missing/interfered tones\commentJ{.}{!} Furthermore, ranging with channel recovery using atomic norm is the second best scheme in terms of performance. We can also conclude that MPS is performing better than WPS, however, they both yield larger ranging error compared to ranging with recovering of missing/interfered tones, especially for \commentJ{small values of Rician factors}{smaller Rician factors}. We also observed that the performance of ranging with zero-padding of the missing/interfered tones is the worst compared other schemes and is very sensitive to the position of tone gaps. This \commentA{highlights}{} the impact of employing proposed schemes, when there are missing/interfered tones.

			\begin{figure*}[t] \centering 
		\includegraphics[scale=0.63]{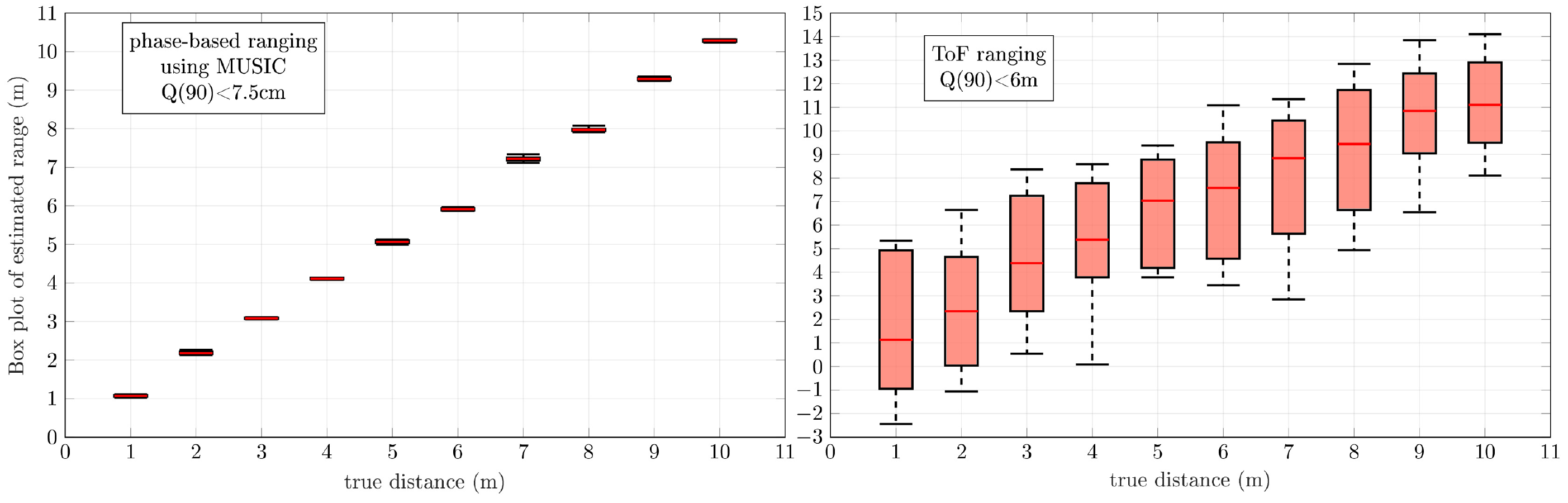}  
		\vspace{-1ex}
		\caption{\R{The box plot of the phase-based ranging using MUSIC (left) and ToF (right) based on the measurement setup of Fig.~\ref{setup_MUSIC_vs_ToF}.}}  \vspace{-2ex}
		\label{MUSIC_vs_ToF} 
	\end{figure*} 

\begin{figure}[!htb]
	\centering
	\includegraphics[trim={2mm 2mm 2mm 2mm},clip,width=.99\columnwidth]{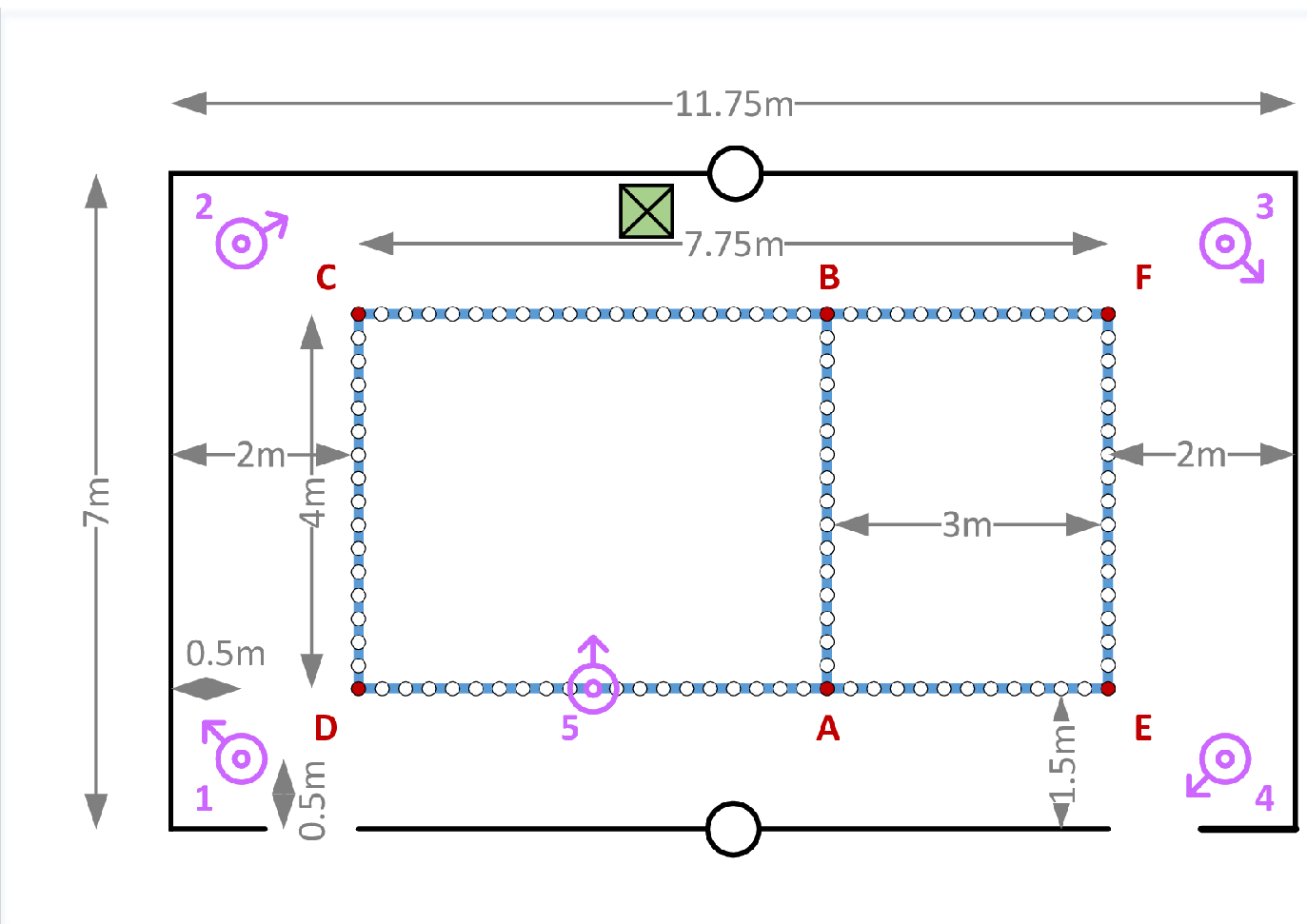}
	\caption{Schematic of measurement setup in meeting room.}
	\vspace{-3ex}
	\label{fig:meas_setup}
\end{figure}

\begin{figure*}[t] \centering 
	\begin{subfigure}{\includegraphics[scale=0.37]{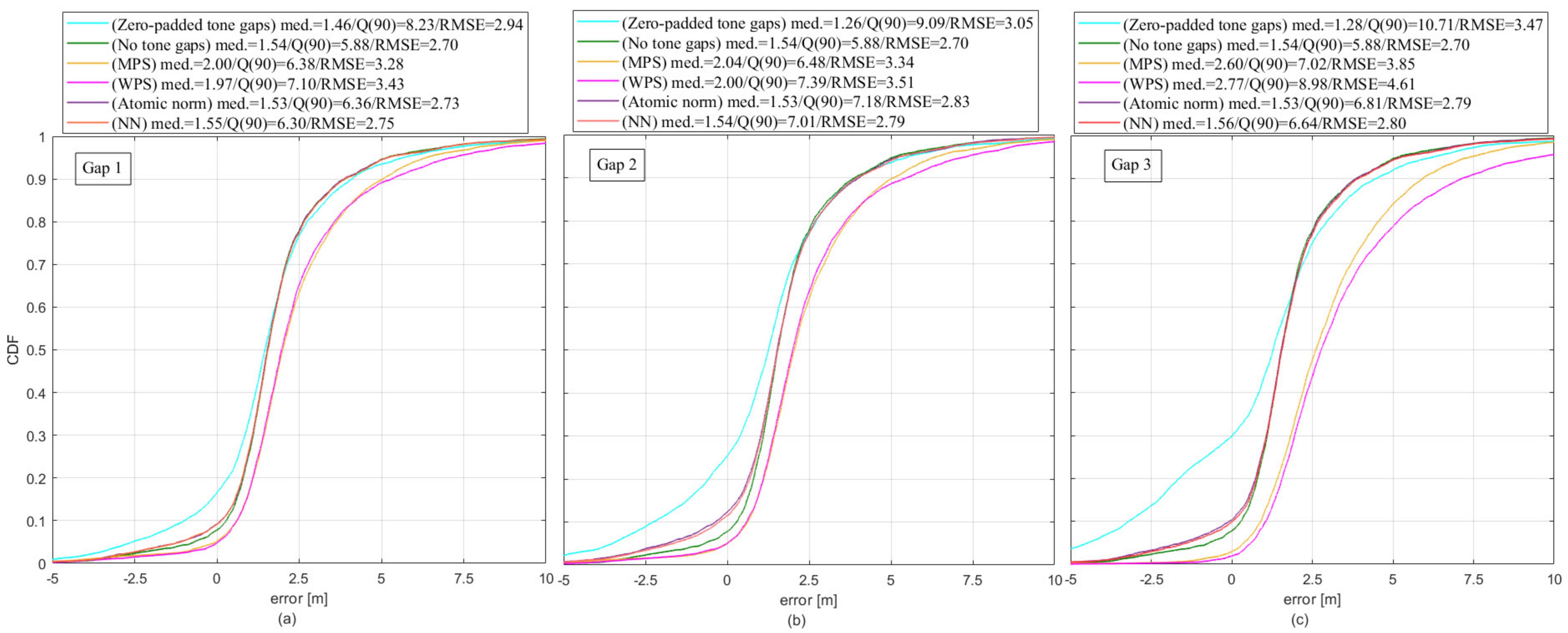}}  
		\vspace{-4ex}
	\end{subfigure}\vfill%
	\begin{subfigure}{\includegraphics[scale=0.37]{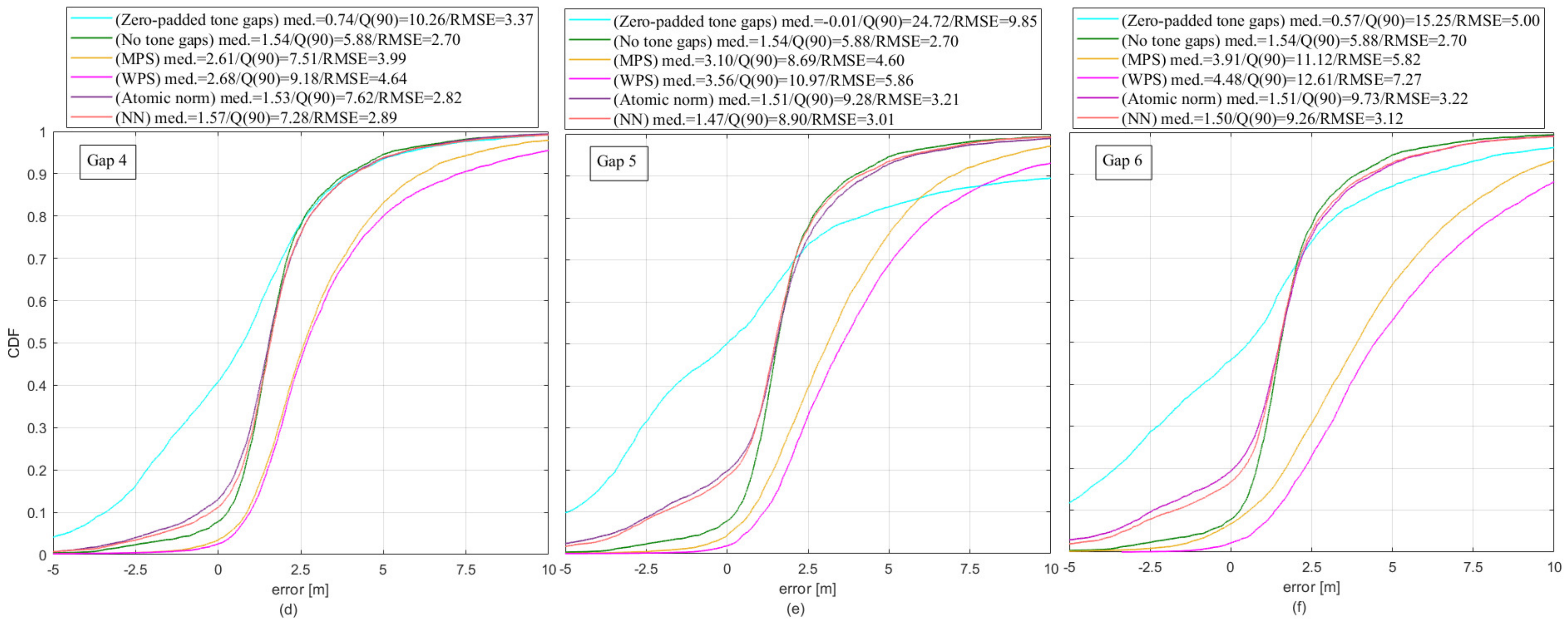} }	
		%	\vspace{-3ex}
		%	\caption{64-QAM} \vspace{1ex}
		%	\label{64_QAM_EDC_singlechannel} 
	\end{subfigure}\vfill%
	\vspace{-2ex}
	\caption{The CDF of range error of different schemes based on post processing on the measured IQ based on the VNA for (a) gap 1, (b) gap 2, (c) gap 3, (d) gap 4, (e) gap 5, (f) gap 6.}  \vspace{-1ex}
	\label{Fig_RMSE1} 
\end{figure*} 

	\begin{figure}[!htb] \centering 
	\includegraphics[scale=0.68]{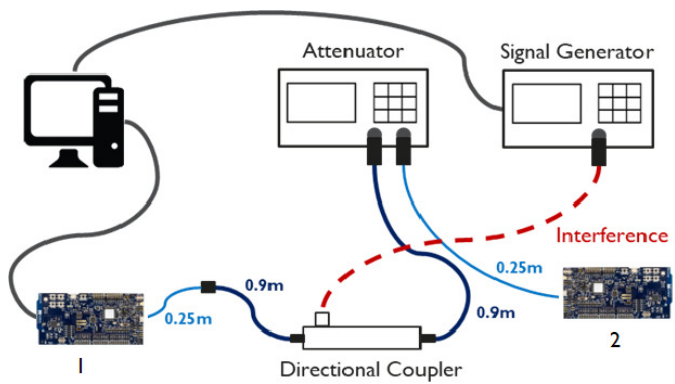}  
	\vspace{-2ex}
	\caption{Schematic of the Measurement setup in Section~\ref{Nordic}.}  \vspace{-3ex}
	\label{Setup1} 
\end{figure} 

	\begin{figure*}[!htb] \centering 
	\includegraphics[scale=0.75]{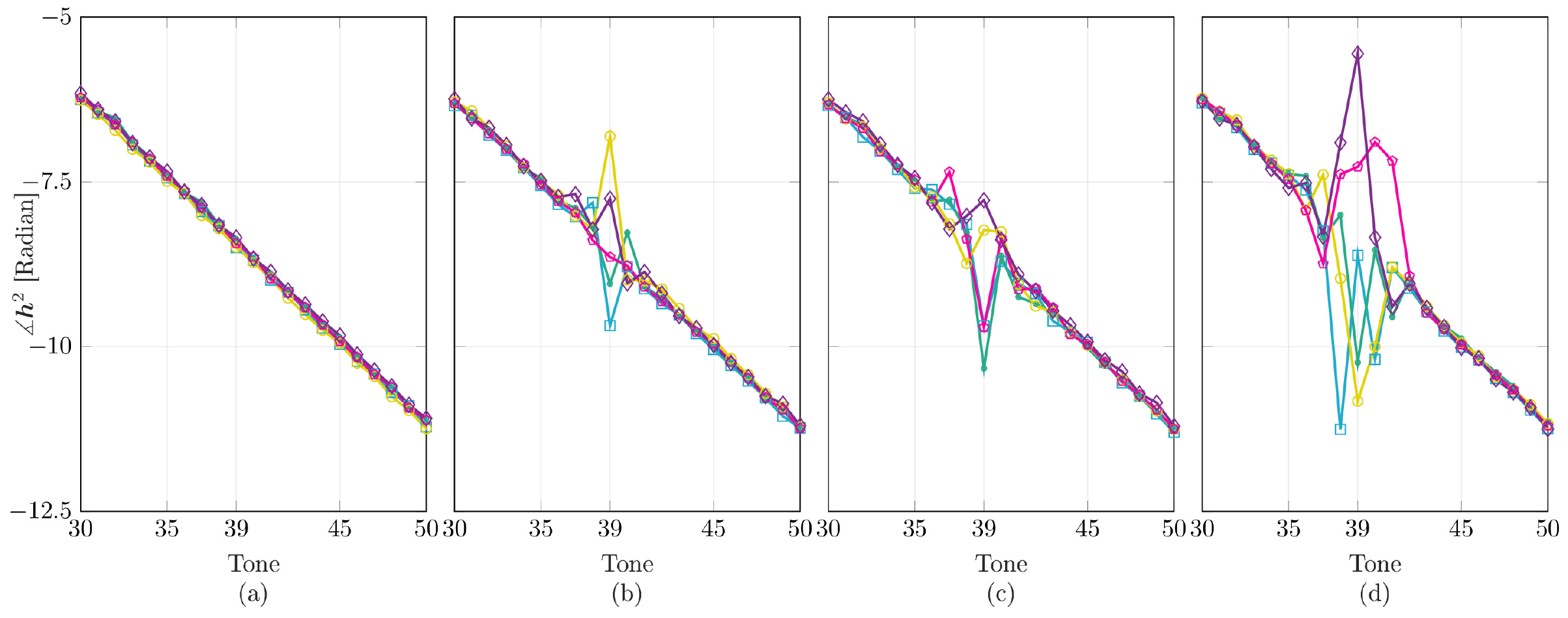}  
	\vspace{-4ex}
	\caption{The progression of the phase of $\measuredangle \boldsymbol{h}^2$ components for tone indices $30$-$50$ for (a) no interference (b) interference with SIR of $0$ dB (c) interference with SIR of $-10$ dB (d) interference with SIR of $-15$ dB.}  \vspace{-2ex}
	\label{Fig_IQ_phase} 
\end{figure*}   
	
	\begin{figure*}[!htb] \centering 
		\includegraphics[scale=0.62]{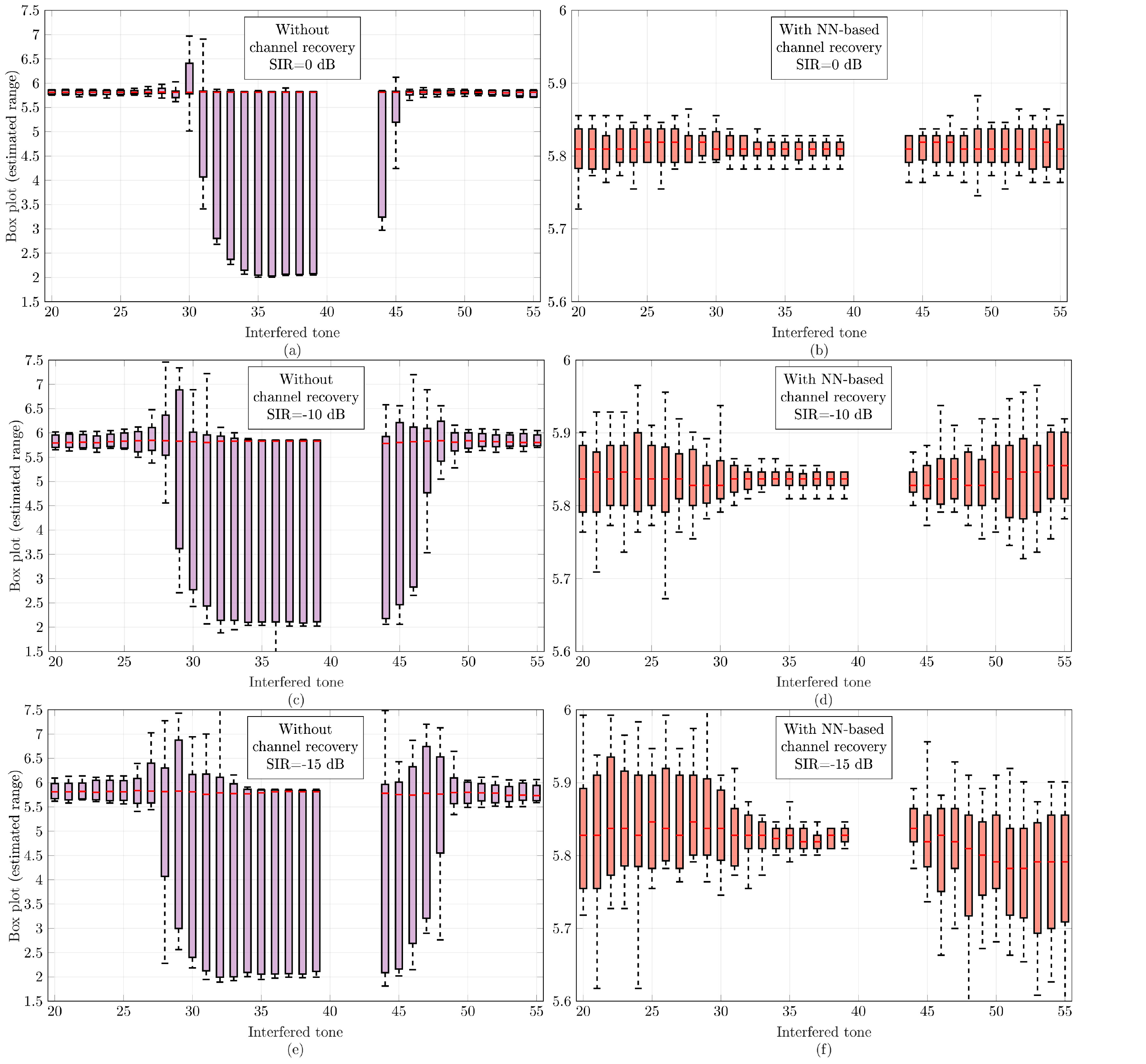}  
		\vspace{-3ex}
		\caption{The box plot of the \RRR{estimated range} of the measurement setup in Fig.~\ref{Setup1} for interfered tone indices of $20-55$. (a), (c), and (e) correspond to ranging system without channel recovery (Fig.~\ref{Ranging_Super}) for SIR of $0$ dB, $-10$ dB, and $-15$ dB, respectively. (b), (d), and (f) ranging system with channel recovery (Fig.~\ref{Ranging_recov}) using NN for SIR of $0$ dB, $-10$ dB, and $-15$ dB, respectively. The values in the Y-axis are in meter.\commentA{}{Super nice if the caption of each subfigure be like "without channel recovery" or "with channel recovery of NN". Besides, I think, we need to remove bias from the boxplots, and other figures. I know that bias can be simply removed, and does not have any impact on ranging precision, but the reviewer might get confused why the error is so large.}}  \vspace{-3ex}
		\label{Box_plots} 
	\end{figure*}

	\subsection{Measurement results}
	
	\subsubsection{\R{Comparing phase-based ranging and ToF in a LoS strong measurement}}
\R{To show the superiority of phase-based ranging based on MUSIC algorithm (see Fig.~\ref{Ranging_Super}) over range estimation using ToF, a wireless outdoor measurements using NXP KW36 (BLE) chip \cite{kw36} are performed. The measurement setup has a strong LoS, as shown in Fig.~\ref{setup_MUSIC_vs_ToF} and is similar to setup considered in \cite{Abidin_El}. The initiator and reflector has omnidirectional antennas. The distance between initiator and reflector are varied from $1$ m to $10$ m with the step size of $1$ m. The total number of measurements per position is $250$ and $K=80$ channels are used for the IQ measurements. The box plot of estimated range for both schemes are shown in Fig.~\ref{MUSIC_vs_ToF}. Each bar in the box plot, shows the median of range error with red line, where the Q90 of ranging is shown with colored bar. One can also see the minimum and maximum of the ranging error of each bar with dashed lines. As it can be seen, both the accuracy and precision of the phase-based ranging using MUSIC is much better than ToF. In particular, for all distances the Q(90) of estimated range improves from $6$ m in ToF to $7.5$ cm in phase-based ranging using MUSIC. In the next measurements setups, we evaluate the effect of channel recovery in phase-based ranging using MUSIC.}

	\subsubsection{Measurement setup based on VNA}
	
	\RRR{The measurements are executed} in a meeting room (without furniture) using a 4-port VNA (Keysight PNA-X N5242A \cite{DIGIKEY2020}), where there are possible interference sources \RRR{such as WIFI-access point} or mobile phones. \RRR{Note that this data set is also used in performance analysis of \cite{boer2019}.} The horizontal and vertical polarized antenna pairs are used for two ports of VNA as an initiator and for the other two ports as the reflector. This leads to $4$ measurements between each antenna combination.  A schematic of the meeting room is depicted in \commentJ{figure}{Fig}~\ref{fig:meas_setup}. The locations of the initiator are shown with numbers $1$-$4$. The location of the reflector \RRR{shown as number $5$}, is located at every $25$cm on the lines between corner points ABCDEF, resulting in $110$ measurement locations. The measurement is done $5$ times on all $K=80$ channels and for the $4$ antenna-pairs. The green box indicates the position of the VNA during the measurements.
	
	In Fig~\ref{Fig_RMSE1}(a)-(f), we show the cumulative distribution function (CDF) of ranging error of the measured data for the tone gaps 1-6, respectively. In each figure, we also show the median, 90-percentage quantile \RRR{Q(90)}, and RMSE of the ranging error. The Q(90) is the ranging error around the median which contains the 90-percentage of the data. Q(90) is a measure for the \commentAFF {accuracy}{precision} of ranging. In general, \RRR{all the conclusions} drawn based on simulated data \commentJ{can be drawn from the CDF of}{apply to} ranging on measured data as well.  Interestingly, similar to Section~\ref{Sec:Sim_r}, the ranging using channel recovery based on NN provides the best compared to other schemes, where its Q(90) \commentA{of}{\it remove} ranging error is up to $35$cm better than the ranging using channel recovery based on atomic norm. Note that the median of range error for both schemes of ranging based on channel recovery is approximately constant ($1.5$ m) for various tone gaps, similar to the reference system of ranging without missing/interfered tone gaps. \RRR{This bias is due to an unknown delay of the antennae in the measurement setup of Fig.~\ref{fig:meas_setup}, as the VNA is calibrated.} This constant error can be removed as bias of range estimation. However, for MPS and WPS, the median of range error is changing more than a meter for various gap, hence, it can not corrected as a bias.
	
	\subsubsection{Measurement setup based on Bluetooth radio nRF52833 board}\label{Nordic}
	
	The schematic of measurement setup is shown in Fig.~\ref{Setup1}. As it can be seen, two "nRF52833" boards \cite{nRF52833}, referred to as board $1$ and board $2$, are connected via an attenuator \cite{siggen} and a directional coupler. The attenuator is tuned to have a received power of $-55$ dBm at both boards. Furthermore, a signal generator \cite{siggen} \commentJ{}{is connected} to the third port of the coupler. Due to the \commentJ{directinality  of}{directionality of the} coupler, the IQ samples at board $1$ are only interfered. In the signal generator, we used Bluetooth Low Energy 1M signal with \RRR{Gaussian frequency shift keying modulation} with continuous packet type, where the \commentJ{}{carrier} frequency of the transmission can be tuned. The power of signal generator is such that the signal-to-interference-ratio (SIR) of $0$ dB, $-10$ dB, and $-15$ dB are achieved. In the measurement, the \commentJ{frequency of interfered tone are}{carrier frequency of interference is} varied from $2.419$ GHz to $2.454$ GHz, corresponding to tone indices of $\{20:55\}$. We remark that due to \commentA{limitations of the application running on the board}{}, it was not possible to capture IQ samples when tones of index $\{40:43\}$ are interfered in our setup.
	
	In Fig.~\ref{Fig_IQ_phase}, the unwrapped phase of the measurement setup for different SIR levels are compared, when the interfered tone is at index $39$. For the sake of comparison, we also show the unwrapped phase of the setup when there is no interference. As one may expect, due to the LoS condition of the setup, \commentJ{we expect the phase varies over a straight line}{the unwrapped phase is a straight-line}, similar to Fig.~\ref{Fig_IQ_phase}(a). However, due to the interference, depending on the interference strength, the phase of tone index $39$ and \RRR{especially tones around it deviate from the straight line}. In other words, the interference, perturbs the phase around interfered tones, which will impact the performance of ranging.  
	
	In Fig.~\ref{Box_plots}, we show the box plot of the \RRR{estimated range} in the measurement setup for different SIR levels, when the position of interfered tone is varied. In this figure, we compare the performance of ranging system without channel recovery (Fig.~\ref{Ranging_Super}) with the performance of ranging system when channel recovery based on NN is employed. Note that the Y-axis of left and right figures are in a different range. As it can be seen, \RRR{the median of range estimate is $5.85$m. This bias is due to cables, connectors, and the attenuator in measurement setup of Fig.~\ref{Setup1}. Furthermore,} the interference especially for the tones close to the middle of the \commentJ{spectrum}{ISM band}, reduces the \RRR{precision} of ranging without employing the channel recovery drastically. \commentJ{This is expected as more components of}{This is expected as a measurement in the middle of the ISM band appears more often in}  the Hankel matrix \RRR{(\eqref{eq:Hankel}, \eqref{eq:Hankel_jthband})}. Hence, the Hankel matrix is changed compared to the case where there is no tone gap. Therefore, the quality of the signal-noise sub-space separation is \commentJ{lower}{reduced}, i.e., there maybe \commentJ{sum}{some} noise sub-spaces with large eigenvalues, which will be \commentJ{considered as}{incorrectly assigned to the} signal \commentJ{}{sub-}space. Furthermore, \commentJ{by}{with} reducing SIR level, the ranging error increases. From Fig.~\ref{Box_plots} it can be inferred that the channel recovery using NN hugely improves the performance of ranging, e.g., for interfered tone of index $39$, the Q90 of ranging error is improved from $4$m to $20$cm. 
	
	We highlight that for the CDF plots shown in Fig.~\ref{Fig_RMSE1} and the box plot in Fig.~\ref{Box_plots}, we employed the NN that is purely trained based on the simulated data, i.e., no re-training based on transfer learning \cite{Pan2010} on the measured data is performed. 
	
	\section{Conclusions}\label{Sec:Conclusions}
	We proposed \commentA{two schemes}{two or three?} to perform ranging, when there are missing/interfered tones in the bandwidth. In \commentJ{}{the} first scheme, the \commentAFF{cost function}{pseudo spectrum} of MUSIC is modified to account for the missing/interfered tones. We \commentJ{}{have} showed that in this scheme MPS outperforms WPS. However, they both perform worse than the second scheme, \commentAFF{}{i.e., ranging based on channel recovery}, which estimates the squared frequency response of missing/interfered tones and then apply the MUSIC to estimate the range. 
	
	\commentA{In the second scheme}{By presenting Example 3 for the second scheme}, we showed that the channel recovery using atomic norm minimization effectively improves the performance at the cost of roughly $3$ orders of magnitude additional computational complexity compared to computational complexity of MUSIC. We further \commentJ{ showed }{have shown} that the computational complexity of channel recovery can be reduced by $4$ orders of magnitude using a trained NN, making its associated additional computational complexity negligible compared to \commentJ{computational complexity of}{} the MUSIC algorithm. We show that the main cost of channel recovery using NN is at most $60.5$ Kbytes of program/flash memory. Interestingly, this scheme improves the RMSE of ranging using channel recovery \commentJ{with}{compared to} atomic norm minimization by up to $35$ cm. Using simulations, we showed that for a small number of missing/interfered tones, ranging with channel recovery using NN performs close to the reference system, where there is no missing/interfered tones. Interestingly, this observation was true for the measured data using the VNA and Nordic platform, which the NN is not trained for. 
	
	Overall, we conclude that the ranging with channel recovery based on NN provides the best performance-complexity trade off, making it an attractive solution for ranging using hardware-limited radios such as BLE.
	
	%Based on the results given in section~\ref{Alg1_des}-\ref{Alg3_des} and the complexity analysis given in section~\ref{Sec:Comp_complexity}, we conclude that the ranging with channel recovery based on NN provides the best performance-complexity trade off, making it an attractive solution for ranging using low-cost radios.

	%\section*{Acknowledgment}
	
	%The authors would like to thank 
	\balance
	%maximizes the possible path delays that can be estimated by finding the MUSIC \comment{Psudo}{Pseudo} spectrum peaks.
	\appendices
	\section{Proof of Preposition 1}\label{APP}
	
	Based on \cite[Theorem 1]{Liao2014MUSICFS}, the possible path delays that can be estimated by finding the MUSIC \commentJ{Psudo}{pseudo} spectrum peaks for Hankel matrix of \eqref{eq:Hankel_jthband} is less than $L_j$ and $b_j-a_j-L_j+2$ values. Therefore, to maximize the number of \commentJ{possible}{resolvable} paths that can be estimated the following optimization should be solved
	\begin{equation}\label{app1}
		\max_{1 \leq L_j \leq b_j-a_j+1} \min (L_j, b_j-a_j-L_j+2),
	\end{equation}  
	\commentJ{Note that intuitively, it is preferable to have the largest number of eigenvalues of \ASH{${\boldsymbol{H}_j^H}\boldsymbol{H}_j$}, as signal-noise sub-space separation is easier.}{such that the rank of ${\boldsymbol{H}_j^H}\boldsymbol{H}_j$ is being maximized.} %One can easily see that \eqref{app1} is equivalent to maximizing the possible Eigen values of ${\boldsymbol{H}_j^H}\boldsymbol{H}_j$, as suggested intuitively.
	
	%The number of nonzero Eigen values of ${\boldsymbol{H}_j^H}\boldsymbol{H}_j$ is equal to the number of nonzero singular values of $\boldsymbol{H}_j$. The number of nonzero singular values of $\boldsymbol{H}_j$ is the rank of the $\boldsymbol{H}_j$ which is always upper bounded by the minimum dimension of $\boldsymbol{H}_j$. Therefore, maximizing the possible number of nonzero Eigen values of ${\boldsymbol{H}_j^H}\boldsymbol{H}_j$ can be written as the following optimization
	To solve \eqref{app1}, we consider two cases: (i) $\min (L_j, b_j-a_j-L_j+2)=L_j$, (ii) $\min (L_j, b_j-a_j-L_j+2)=b_j-a_j-L_j+2$. In case (i), \eqref{app1} boils down to $\max_{1 \leq L_j \leq \lfloor \frac{{b_j}-{a_j}+1}{2}\rfloor +1} L_j$, which has the obvious solution of $L_j=\lfloor \frac{{b_j}-{a_j}+1}{2}\rfloor +1$. For the case (ii), \eqref{app1} boils down to $\max_{\lfloor \frac{{b_j}-{a_j}+1}{2}\rfloor +1 \leq L_j \leq b_j-a_j+1}  b_j-a_j-L_j+2$. The objective function in this case maximize for the minimum values of $L_j$, hence, the solution of case (ii) is $L_j=\lfloor \frac{{b_j}-{a_j}+1}{2}\rfloor +1$. As in both cases, the solution in similar, the solution of \eqref{app1} is also $L_j=\lfloor \frac{{b_j}-{a_j}+1}{2}\rfloor +1$.
	
	%\bibliographystyle{IEEEtran}
	%\bibliography{IEEEabrv,book}
	

\end{document}